%% file: main.tex
\documentclass[letterpaper,prc,twocolumn,showpacs,floatfix,nofootinbib,preprintnumbers,superscriptaddress,amsmath,amssymb]{revtex4-2}
\usepackage{CJK}
\usepackage{hyperref}
\usepackage{amsmath}
\usepackage{amsfonts}
\usepackage{physics}
\usepackage{xspace}
\usepackage{graphicx}
\usepackage[dvipsnames]{xcolor}
\usepackage{dcolumn}           
\usepackage{bm}
\usepackage{mathtools}
\usepackage{algorithm}
\usepackage{soul}
\usepackage[noend]{algpseudocode}
\usepackage{bbm}
\usepackage[T1]{fontenc}
\usepackage[rightcaption]{sidecap}
\sidecaptionvpos{figure}{c}
\allowdisplaybreaks

\newcommand{\photo}{\sigma_{\rm abs}}
\renewcommand{\vec}{\boldsymbol}
\newcommand{\gsf}{$\gamma$SF}
\newcommand{\enm}{E^\text{NM}/A}

\newcommand{\rhonm}{\rho_{\rm c}}
\newcommand{\knm}{K^\text{NM}}
\newcommand{\ms}{M_{s}^*}

\newcommand{\asym}{a_\text{sym}^\text{NM}}

\newcolumntype{z}[1]{D{.}{.}{#1}}

\begin{document}

\preprint{LLNL-JRNL-866328}
\preprint{LA-UR-24-27051}

\begin{CJK*}{UTF8}{gbsn}

\title{Multipole responses in fissioning nuclei and their uncertainties}

\author{Tong \surname{Li} (李通)}
\email[]{li94@llnl.gov}
\affiliation{Nuclear and Chemical Sciences Division, Lawrence Livermore National Laboratory, 
Livermore, CA 94551, USA}

\author{Nicolas \surname{Schunck}}
\email[]{schunck1@llnl.gov}
\affiliation{Nuclear and Chemical Sciences Division, Lawrence Livermore National Laboratory, 
Livermore, CA 94551, USA}

\author{Mike \surname{Grosskopf}}
\email[]{mikegros@lanl.gov}
\affiliation{Computer, Computational, and Statistical Sciences Division, Los Alamos National Laboratory, 
Los Alamos, NM 87545, USA}

\begin{abstract}
Electromagnetic multipole responses are key inputs to model the structure, 
decay and reaction of atomic nuclei. With the introduction of the finite
amplitude method (FAM), large-scale calculations of the nuclear linear response 
in heavy deformed nuclei have become possible. This work provides a detailed
study of multipole responses in actinide nuclei with Skyrme energy density
functionals. We quantify both systematic and statistical uncertainties
induced by the functional parameterization in FAM calculations. 
We also extend the FAM formalism to perform blocking calculations 
with the equal filling approximation for odd-mass and odd-odd nuclei, 
and analyze the impact of blocking configurations on the 
response. By examining the entire plutonium isotopic chain from the proton to 
the neutron dripline, we find a large variability of the response with the neutron 
number and study how it correlates with the deformation of the nuclear 
ground state.
\end{abstract}

\maketitle

\end{CJK*}


\section{Introduction}
\label{sec:intro}

Photonuclear processes are amongst the most valuable probes into the structure 
and reactions of atomic nuclei. Electromagnetic transitions between discrete 
low-lying excited states are used to reconstruct the excitation spectrum of the 
nucleus with the help of coincidence techniques \cite{gilmore2008practical,
dunn2021gammaray}. At higher energies, the emission of photons becomes 
statistical and is typically quantified by the $\gamma$-strength function (\gsf) 
\cite{bartholomew1973gamma}. Invoking the Brink-Axel hypothesis 
\cite{brink1955thesis,axel1971hypothesis} allows using these \gsf{s} to 
characterize the absorption of a photon by the nucleus in statistical reaction 
theory. As such, they play an essential role both in fundamental physics such 
as nucleosynthesis, or in applications ranging from medical isotope 
production to fission technology and nuclear forensics.

Discrete electromagnetic transitions can be computed directly as the matrix 
elements of relevant operators -- the multipole operators of the 
electromagnetic field -- between initial and final nuclear states. 
Configuration-interaction (CI) methods such as the nuclear shell model are well 
suited to this task \cite{caurier2005shell}. The shell model can also be used 
to compute \gsf{s} in selected mass regions thanks to averaging procedures 
\cite{stetcu2003tests,sieja2017electric,sieja2018shellmodel}. In heavy deformed 
nuclei, direct CI methods often become impractical and microscopic calculations 
of electromagnetic transitions are performed with collective models such as the 
Bohr Hamiltonian \cite{prochniak2009quadrupole,delaroche2010structure} or 
within symmetry-conserving, multi-reference energy density functional theory; 
see, e.g., \cite{bender2008configuration,niksic2009relativistic,egido200410,
bally2014meanfield,egido2016stateoftheart,borrajo2018symmetry} for a selection
of relatively recent applications. To compute \gsf{s}, the linear response
theory based on the quasiparticle random phase approximation (QRPA) remains by
far the most common approach; see, e.g., 
\cite{goriely2002largescale,paar2003quasiparticle,goriely2004microscopic,
yuksel2017multipole,hilaire2017quasiparticle,goriely2018gognyhfb,
xu2021systematical,kaur2024electric} for examples of such studies with 
Skyrme, Gogny, and covariant energy functionals. 
Theories going beyond the linear response 
have only been applied on a case-by-case basis as they are computationally a 
lot more expensive \cite{litvinova2013relativistic,papakonstantinou2014second,
gambacurta2015subtraction,litvinova2015nuclear,gambacurta2016second,
litvinova2019toward,litvinova2022microscopic,litvinova2023relativistic,
yang2024magnetic}.

For a long time, fully self-consistent QRPA calculations in heavy deformed 
nuclei were hampered by their large computational costs 
\cite{terasaki2010selfconsistent,terasaki2011testing,martini2016largescale}.
However, the invention of the finite-amplitude method (FAM) made calculations of the 
linear response in such nuclei much more accessible \cite{nakatsukasa2007finite,
avogadro2011finite,inakura2009selfconsistent}. The focus of early studies on the FAM 
was largely to establish benchmarks against the direct QRPA 
\cite{stoitsov2011monopole,hinohara2013lowenergy,niksic2013implementation,
hinohara2015complexenergy} and probe its feasibility and predictive power 
for multipole responses or photoabsorption cross sections 
\cite{kortelainen2015multipole,oishi2016finite}.
The goal of this paper is to apply the finite amplitude method to perform a 
more systematic analysis of the multiple response in actinide nuclei. In 
particular, we pay special attention to estimating theoretical uncertainties 
originating from the parameterization of the energy functional and test the 
extension of the FAM for odd-mass and odd-odd nuclei. This work is a 
prelude to a more large-scale calculation of \gsf{s} at the scale of the entire 
mass table.

This paper is organized as follows. 
In Sec.~\ref{sec:theory} we briefly summarize the like-particle FAM formalism 
for electromagnetic operators. The numerical setup of our 
calculations is presented in Sec.~\ref{sec:numerics}. Benchmark calculations 
of photoabsorption cross sections and electromagnetic multipole responses 
are shown in Sec.~\ref{sec:responses_even_even} with detailed analyses of their uncertainties.
Properties of electromagnetic responses in actinides nuclei are then discussed in Sec.~\ref{sec:actinides}. 
Additional technical details, such as the quasiparticle cutoff procedure used in the calculation of 
transition densities and equations for energy-weighted sum rules, 
are given in the Appendices; 
additional figures and tables are provided in the Supplemental Material \cite{supp}. 


\section{Theoretical Framework}
\label{sec:theory}

The FAM formalism for an even-even nucleus was presented in 
Refs.~\cite{nakatsukasa2007finite, avogadro2011finite}, and its extension to 
odd-$A$ and odd-odd nuclei for $\beta$-decay calculations was published in 
Ref.~\cite{shafer2016beta}, followed by Ref.~\cite{giraud2022finitetemperature} 
for the applications of the finite-temperature (FT) FAM on electron capture. In 
Sec.~\ref{subsec:fam} we give a unified summary of the FAM that incorporates 
the cases of even-even, odd-$A$ and odd-odd nuclei, as well as the FT system. 
In Sec.~\ref{subsec:ext_fields}, we provide formulas relevant to the 
application of the FAM to electromagnetic responses. 


\subsection{Finite amplitude method}
\label{subsec:fam}

In the FAM, an external perturbation is applied on a nucleus to induce 
oscillations around a static Hartree-Fock-Bogoliubov (HFB) state, which can be 
described by the small-amplitude limit of the time-dependent HFB (TDHFB) 
equation. The HFB state of an even-even nucleus is a vacuum with respect to 
Bogoliubov quasiparticle operators, i.e., $\beta_\mu |\Phi\rangle=0$. To 
compute an odd-$A$ nucleus, we employ the equal filling approximation (EFA) 
\cite{duguet2001pairing,perez-martin2008microscopic,schunck2010onequasiparticle} 
to preserve the time-reversal symmetry in the static HFB calculation. 
In the EFA, all the densities are computed by 
averaging the densities of the one-quasiparticle state 
$\beta^\dagger_{\mu_B} |\Phi\rangle$ and its time-reversed partner 
$\beta^\dagger_{\bar{\mu}_B} |\Phi\rangle$. An odd-odd nucleus can be similarly 
described by averaging the densities of a two-quasiparticle state and 
its time-reversed partner. The EFA-HFB state can be treated as a special case 
of the FT-HFB state \cite{perez-martin2008microscopic}. Both of them can be 
described as a statistical ensemble whose density operator is 
\begin{equation}\label{eq:density_operator_HFB}
    \begin{aligned}
        \hat{\mathcal{D}}=& |\Phi\rangle\langle\Phi|+\sum_\mu \beta_\mu^{\dagger}| \Phi\rangle p_\mu\langle\Phi| \beta_\mu \\
        &+\frac{1}{2!} \sum_{\mu \nu} \beta_\mu^{\dagger} \beta_\nu^{\dagger}|\Phi\rangle p_\mu p_\nu\langle\Phi| \beta_\nu \beta_\mu+\cdots ,
    \end{aligned}
\end{equation}
where $p_\mu = \frac{f_\mu}{1-f_\mu}$ and $f_\mu$ is the quasiparticle occupancy. 
In the EFA the occupancy $f_\mu$ is 
\begin{equation}\label{eq:EFA_occupation}
    f_\mu= 
    \begin{cases}
        \frac{1}{2} & \mu = \mu_B \text{ or } \bar{\mu}_B , \\
        0 & \text{otherwise}, 
    \end{cases}
\end{equation}
while in the FT-HFB calculation it is given by the Fermi-Dirac distribution 
\begin{equation}
    f_\mu = \frac{1}{e^{E_\mu/(k_B T)}+1},
\end{equation}
where $E_\mu$ is the quasiparticle energy and $T$ is the temperature. The 
corresponding generalized density matrix is 
\begin{equation}\label{eq:R_qp_basis}
    \mathbb{R}_{\mu\nu} = 
    \begin{pmatrix}
        \langle \beta^\dagger_\nu \beta_\mu \rangle & \langle  \beta_\nu \beta_\mu \rangle \\
        \langle  \beta^\dagger_\nu \beta^\dagger_\mu \rangle & \langle \beta_\nu \beta^\dagger_\mu \rangle
    \end{pmatrix} =
    \begin{pmatrix}
        f_\mu \delta_{\mu\nu}  & 0 \\
        0 & 1- f_\mu \delta_{\mu\nu}
    \end{pmatrix}. 
\end{equation}
We see that an even-even nucleus can also be treated as a statistical ensemble 
with $f_\mu=0$ for all the quasiparticles. It should be noted that all the 
discussions above apply to both static and time-dependent HFB states. 

In the FAM we assume that the nucleus oscillates under a weak external field of 
frequency $\omega$: 
\begin{equation}\label{eq:F_oscillate}
    F(t) =\eta\Big[ F(\omega) e^{-i \omega t}+F^{\dagger}(\omega) e^{i \omega t} \Big],
\end{equation}
where $\eta$ is a small real number and 
\begin{equation}
    F(\omega) =\frac{1}{2} \sum_{\mu\nu}\left(F_{\mu\nu}^{20} A_{\mu\nu}^{\dagger}+F_{\mu\nu}^{02} A_{\mu\nu}\right)
    +\sum_{\mu\nu} F_{\mu\nu}^{11} B_{\mu\nu},
\end{equation}
where $A^\dagger_{\mu\nu} \equiv \beta^\dagger_\mu \beta^\dagger_\nu$,  
$B_{\mu\nu} \equiv \beta^\dagger_\mu \beta_\nu$, and $F^{20}$ and $F^{02}$ are 
antisymmetric matrices. As shown in \cite{nakatsukasa2012density}, taking into 
account this time-dependent operator into the TDHFB equation results in small 
oscillations of the TDHFB mean field:
\begin{subequations}\label{eq:TDHFB_H_decompose}
    \begin{align}
        \mathcal{H}(t) &= \mathcal{H}_0 + \eta\Big[ \delta\mathcal{H}(\omega) e^{-i \omega t} + \delta\mathcal{H}^{\dagger}(\omega) e^{i \omega t} \Big], \\
        \delta\mathcal{H}(\omega) &=\frac{1}{2} \sum_{\mu\nu}\left( \delta H_{\mu\nu}^{20} A_{\mu\nu}^{\dagger} + \delta H_{\mu\nu}^{02} A_{\mu\nu} \right)
        + \delta H_{\mu\nu}^{11} B_{\mu\nu},
    \end{align}
\end{subequations}
where $\mathcal{H}_0 = \sum_\mu E_\mu B_{\mu\mu}$ is the static HFB Hamiltonian.
The time-dependent quasiparticle operator can be decomposed in a similar manner 
as 
\begin{equation}\label{eq:TD_qp_op}
    \beta_{\mu}(t)
    =\Big[ \beta_{\mu}+\eta \delta \beta_{\mu}(t) \Big] e^{i E_{\mu} t},
\end{equation}
where $\beta_\mu$ is the quasiparticle operator of the static HFB solution, and 
$\delta\beta_\mu(t)$ can be written as 
\begin{equation}\label{eq:XYPQ_def}
    \begin{aligned}
        \delta \beta_{\mu}(t) =&
        \sum_{\nu} \beta_{\nu}^{\dagger}\Big[ X_{\nu\mu}(\omega) e^{-i \omega t}+Y_{\nu\mu}^{*}(\omega) e^{i \omega t} \Big] \\
        &+ \sum_{\nu} \beta_\nu \Big[ P_{\mu\nu}(\omega) e^{-i \omega t} + Q^*_{\mu\nu}(\omega) e^{i \omega t} \Big],
    \end{aligned}
\end{equation}
where $X$, $Y$, $P$ and $Q$ are called FAM amplitudes. Using the unitarity 
of the Bogoliubov transformation, we can prove $Q = -P^T$ while $X$ and $Y$ are 
antisymmetric; see Appendix \ref{app:FAM_amplitudes} for details. 

We can substitute the expressions of $F(t)$, $\mathcal{H}(t)$ and 
$\beta_\mu(t)$ into the TDHFB equation to get
\begin{equation}\label{eq:TDHFB}
    i \frac{\partial \beta_{\mu}(t)}{\partial t}=\left[\mathcal{H}(t)+F(t), \beta_{\mu}(t)\right],
\end{equation}
where we use $\hbar=1$. 
Expanding Eq.~\eqref{eq:TDHFB} up to the first order in $\eta$, we obtain the FAM 
equations
\begin{subequations}\label{eq:FAM_equation}
    \begin{align}
        \label{eq:FAM_equation_P}
        \left(E_\mu - E_\nu -\omega\right) P_{\mu\nu}(\omega) - \delta H^{11}_{\mu\nu}(\omega) - F^{11}_{\mu\nu}&=0, \\
        \left(E_\mu + E_\nu -\omega\right) X_{\mu\nu}(\omega) + \delta H^{20}_{\mu\nu}(\omega) + F^{20}_{\mu\nu}&=0, \\
        \left(E_\mu + E_\nu +\omega\right) Y_{\mu\nu}(\omega) + \delta H^{02}_{\mu\nu}(\omega) + F^{02}_{\mu\nu}&=0. 
    \end{align}
\end{subequations}
Expanding the induced mean field $\delta H$ in terms of the FAM amplitudes $(P,X,Y)$  
yields the FT-QRPA equation presented in Ref.~\cite{sommermann1983microscopic}.

Constructing the matrix involved in the FT-QRPA equation is numerically very 
expensive since it has a dimension of $4\times N_{2\mathrm{qp}}$, where 
$N_{2\mathrm{qp}}$ is the number of two-quasiparticle excitations: in heavy 
nuclei where there could be about $10^3$ relevant quasiparticles, this leads to 
a QRPA matrix with a dimension of order $10^{6}$, that is, about $10^{12}$ matrix 
elements to compute. In the FAM, we calculate the induced densities listed in 
Appendix \ref{app:induced_quantities} from the FAM amplitudes to determine the 
induced mean fields \eqref{eq:TDHFB_H_decompose}, which allows solving the FAM 
equations \eqref{eq:FAM_equation}. This procedure is iterative but avoids 
explicitly building the FT-QRPA matrix, which significantly accelerates 
the calculation. In Appendix \ref{app:induced_quantities} we provide more details 
about the induced densities and mean fields, and show why the amplitude $P$ does 
not contribute in an even-even nucleus. We also discuss how to implement the 
quasiparticle cutoff to avoid the ultra-violet divergence brought by zero-range 
pairing interactions \cite{bulgac2002renormalization,borycki2006pairing,
schunck2019energy}. 


\subsection{Transition strength and electromagnetic transition operators}
\label{subsec:ext_fields}

Using the solutions of Eq.~\eqref{eq:FAM_equation}, we can compute the transition strength distribution 
\cite{nakatsukasa2007finite, avogadro2011finite}
\begin{equation}\label{eq:transition_strength}
    \frac{d B(\omega; F)}{d \omega} = \sum_{n>0}|\langle n |F| 0 \rangle|^{2} \delta\left(\omega-\Omega_{n}\right) 
    =-\frac{1}{\pi} \operatorname{Im} S(\omega; F), 
\end{equation}
where $\Omega_n$ is the excitation energy of $|n\rangle$ and $S(\omega; F)$ is 
the FAM response function given by 
\begin{equation}\label{eq:FAM_response}
    S(\omega; F) = \mathrm{Tr}\left( f^\dagger \delta\rho \right) = \sum_{kl} f^*_{kl} \delta\rho_{kl}(\omega; F),
\end{equation}
where $f_{kl}$ are the matrix elements of the external field in the 
single-particle (s.p.) basis ($F(\omega) = \sum_{kl} f_{kl} c^\dagger_k c_l$), 
and the induced density matrix $\delta\rho$ is given by 
Eq.~\eqref{eq:rho_kappa_from_PQXY_explicit}. In practical calculations, we 
usually adopt a complex frequency $\omega \rightarrow \omega+i\gamma$, and the 
corresponding strength distribution is smeared by a Lorentzian function 
with a half width $\gamma$ (full width $\Gamma=2\gamma$): 
\begin{equation}\label{eq:smeared_transition_strength}
    \frac{d B}{d \omega} = 
    \frac{\gamma}{\pi} \sum_{n}\left\{\frac{|\langle n|F|0 \rangle|^{2}}{\left(\omega-\Omega_{n}\right)^2+\gamma^2}
    -\frac{|\langle n|F^{\dagger}|0 \rangle|^{2}}{\left(\omega+\Omega_{n}\right)^2+\gamma^2}\right\}. 
\end{equation}
The smearing width $\gamma$ mainly accounts for the spreading effect brought by mode-mode 
coupling \cite{harakeh2001giant}. In linear response theory where such 
coupling effects are not computed, the smearing width should in principle be adjusted 
to match experimental evaluations \cite{yoshida2011dipole}. In this paper we 
adopt a standard width of $\Gamma=1$ MeV, unless otherwise stated, and perform FAM 
calculations on a $\omega$ grid with 0.1 MeV spacing, which is fine enough for 
$\Gamma=1$ MeV.

In this work we study electromagnetic multipole transitions in heavy nuclei. 
The electric multipole transition operator is written in spherical coordinates 
as \cite{ring2004nuclear}
\begin{equation}\label{eq:electric_multipole_op}
    Q_{L K}  = e \sum_{i=1}^A q_{\tau_i} r_i^L Y_{L K}(\Omega_i), 
\end{equation}
where particle $i$ is either a neutron ($\tau_i = n$) or a proton ($\tau_i = p$), 
$q_{\tau}$ is the effective charge in the unit of the elementary charge $e$, 
$\Omega_i \equiv (\theta_i,\varphi_i)$ are the angular coordinates, and 
$Y_{LK}(\Omega_i)$ the spherical harmonics \cite{varshalovich1988quantum}. 
The magnetic multipole operator is \cite{ring2004nuclear}
\begin{equation}
    M_{L K}  = \mu_\mathrm{N} \sum_{i=1}^A 
    \vec{\nabla} \left[r_i^L Y_{L K}(\Omega_i) \right] \cdot
    \left( \frac{2g_{l}^{(\tau_i)}}{L+1} \vec{l}_i + g_{s}^{(\tau_i)}\vec{s}_i \right), 
\end{equation}
where $\mu_N$ is the nuclear magneton, 
$\vec{l}_i = \left(\vec{r}_i \times \vec{p}_i\right) / \hbar$ the orbital 
angular-momentum operator, $\vec{s}_i = \frac{1}{2} \vec{\sigma_i}$ the spin 
operator, and $g_{l}^{(\tau)}$ and $g_{s}^{(\tau)}$ are the effective $g$ 
factors of a nucleon of type $\tau$. In this paper we use the bare charges and 
bare $g$ factors of nucleons in multipole operators $Q$ and $M$, i.e., we take 
$q_n = 0$, $q_p = 1$, $g_l^{(n)}=0$, $g_l^{(p)}=1$, $g_s^{(n)}=5.586$, and 
$g_s^{(p)}=-3.826$ \cite{tiesinga2021codata}. It has been suggested that the 
spin $g$ factors be quenched to match the calculations of $M1$ responses with 
experimental data \cite{harakeh2001giant}; this problem is beyond the scope of 
this paper. In addition, we decompose the multipole operator $V=Q$ or $M$ into 
isoscalar (IS) and isovector (IV) components as 
\begin{equation}
    V = \frac{1}{2} \left( V^{\mathrm{IS}} + V^{\mathrm{IV}} \right), 
\end{equation}
where neutrons and protons take equal (opposite) effective charges or $g$ 
factors in $V^{\mathrm{IS}}$ ($V^{\mathrm{IV}}$). 

For the electric dipole ($E1$) operator, we can explicitly remove the 
center-of-mass motion and obtain following effective charges 
\cite{ring2004nuclear}
\begin{equation}\label{eq:E1_charges_cm}
    q_n = -\frac{Z}{A}, \quad q_p = \frac{N}{A}. 
\end{equation}
One can see that when $N=Z$ the $E1$ operator becomes purely isovector. When 
$N$ and $Z$ do not differ much, the $E1$ operator is mostly isovector and the 
corresponding response will heavily depend on the isovector component of the 
energy density functional (EDF). 

With the $E1$ operator, we can calculate the cross section for the absorption 
of dipole radiation. Assuming that incoming photons with energy $\omega$ come 
along the $z$ axis in the lab frame, the $E1$ photoabsorption cross section is 
\cite{ring2004nuclear} 
\begin{equation}\label{eq:photoabsorption_cross_section}
        \photo(\omega) = 
        \frac{16\pi^3}{3} \alpha \sum_{n>0} \omega 
        \left| \langle \Psi_n | Q_{10}^{\mathrm{eff}} | \Psi_0 \rangle \right|^2 \delta(\omega - \Omega_n),
\end{equation}
where $\alpha = \frac{e^2}{\hbar c}$ is the fine structure constant, 
$|\Psi_0\rangle$ is the ground state and $|\Psi_n\rangle$ the $n^{\rm th}$ 
excited state, both in the laboratory frame, and $\Omega_n$ is the excitation 
energy. The superscript ``eff'' on $Q_{1K}$ denotes the use of effective charges 
given in Eq.~\eqref{eq:E1_charges_cm}. In this work we ignore the contributions 
from magnetic and high-order electric transitions to the photoabsorption cross 
section, since they are usually much smaller than that from the $E1$ 
transition \cite{lipparini1989sum}. 

For a deformed nucleus, both HFB and FAM calculations are carried out in the 
intrinsic frame of the nucleus, so we need to perform angular-momentum 
projection (spherical symmetry restoration) \cite{ring2004nuclear,peru2014mean,
chimanski2023addressing} to obtain wavefunctions in the laboratory frame. After 
projection, the $E1$ photoabsorption cross section 
\eqref{eq:photoabsorption_cross_section} can be expressed as 
\cite{inakura2009selfconsistent,yoshida2011dipole,oishi2016finite}
\begin{equation}\label{eq:photoabsorption_cross_section_intrinsic}
    \photo(\omega) = \frac{16\pi^3}{9} \alpha \sum_{K=0, \pm1}\frac{dB(\omega; Q_{1K}^{\mathrm{eff}})}{d\omega}, 
\end{equation}
where $dB/d\omega$ is evaluated in the intrinsic frame via 
Eqs.~\eqref{eq:transition_strength} and \eqref{eq:FAM_response}. One can find 
more details in Ref.~\cite{chimanski2023addressing} on the angular-momentum 
projection in QRPA and FAM calculations. 

When the static HFB solution is time-reversal invariant, the FAM responses of 
multipole operators $V_{LK}$ and $V_{L -K}$ are equal: only the results of 
$K=0$ and $1$ in Eq.~\eqref{eq:photoabsorption_cross_section_intrinsic} need to 
be calculated. When the static HFB solution has spherical symmetry, the FAM 
response will not depend on $K$, and we can thus further reduce the 
computational cost.


\section{Numerical Setup}
\label{sec:numerics}

With the formalism presented in Sec.~\ref{sec:theory}, we have developed a new 
numerical program called \textsc{gfam} to perform axially-deformed FAM 
calculations with Skyrme EDFs for nuclear 
electromagnetic transitions. \textsc{gfam} is based on the charge-changing FAM 
code \textsc{pnfam} \cite{mustonen2014finiteamplitude,shafer2016beta,
giraud2022finitetemperature}, which was successfully employed to conduct global 
calculations on the $\beta$ decay of even-even, odd-$A$ and odd-odd nuclei 
\cite{mustonen2016global,ney2020global}, as well as the finite-temperature 
electron capture \cite{giraud2022finitetemperature}. Like \textsc{pnfam}, the 
\textsc{gfam} program is also computationally efficient and amenable to 
large-scale studies. One of its inputs is the static HFB solution from the 
\textsc{hfbtho} code \cite{stoitsov2005axially,stoitsov2013axially,
perez2017axially,marevic2022axiallydeformed}.
Parity is not necessarily conserved in the HFB state, 
but we will preserve it to reduce computational costs unless otherwise stated. 

In our HFB and FAM calculations, all the quasiparticle wavefunctions are 
expanded in a deformed harmonic-oscillator (HO) basis up to $N_0 = 20$ 
shells. The oscillator length is given by the default setting in \textsc{hfbtho} 
\cite{stoitsov2013axially}. We also vary the quadrupole deformation $\beta_2$ 
of the HO basis and initial constraint between -0.2 and 0.2 to locate the HFB 
ground state. In the particle-particle channel we use a mixed surface-volume delta force, 
with a pairing cutoff of $E_{\rm cut}=60$ MeV \cite{dobaczewski2002contact}.
With the numerical parameters given above, it takes less than an hour on a node 
with 56 CPU cores to solve the FAM equations over 500 $\omega$ points, which is 
fast enough for future surveys over the nuclear chart. One can find discussions 
in Ref.~\cite{li2024numerical} about the dependence of FAM responses on the 
numerical parameters mentioned above. 

In this work, we consider several Skyrme EDFs fitted under various protocols
to show the systematic uncertainties of FAM results. The SLy4 (without tensor 
terms) and SLy5 (with tensor terms) functionals were fitted to improve the 
description of neutron-rich nuclei and included a constraint on the 
Thomas-Reiche-Kuhn (TRK) sum rule \cite{chabanat1998skyrme}. The SkM* parametrization 
is the reference EDF for the studies of deformed nuclei, especially nuclear fission 
\cite{bartel1982better}. The SkI3 functional was fitted largely to improve the 
description of the isotope shifts in the Pb region \cite{reinhard1995nuclear}.
Finally, the HFB1 (UNEDF1-HFB) functional is the only one in our set where the 
particle-particle channel is fitted simultaneously with the particle-hole channel 
\cite{schunck2015error}. It is similar to the UNEDF1 functional \cite{kortelainen2012nuclear}, only without 
the Lipkin-Nogami prescription for pairing. These EDFs predict a relativey broad 
range of nuclear matter properties as shown in Table \ref{tab:nm}, where
we follow the notations of Ref.~\cite{kortelainen2010nuclear}. 

\begin{table}[!htb]
\caption{Summary of nuclear matter properties for the five energy 
functionals considered in this work. The saturation density $\rhonm$ is in 
fm$^{-3}$; the binding energy per nucleon $\enm$, incompressibility $\knm$,
symmetry energy $\asym$ are in MeV; the isoscalar effective mass $\ms$ and 
enhancement factor of the Thomas-Reiche-Kuhn sum rule $\kappa_{\rm TRK}$ \eqref{eq:kappa_NM} 
are dimensionless.}
\label{tab:nm}
\begin{ruledtabular}
\begin{tabular}{ccccccc}
     & $\rhonm$ & $\enm$ & $\knm$ & $\ms$ & $\asym$ & $\kappa_{\rm TRK}$ \\
\hline
SLy4 & 0.160 & -15.97 & 229.90 & 0.694 & 32.00 & 0.250 \\
SLy5 & 0.160 & -15.98 & 229.91 & 0.698 & 32.01 & 0.250 \\
SkM* & 0.160 & -15.77 & 216.66 & 0.786 & 30.03 & 0.532 \\
SkI3 & 0.158 & -15.96 & 257.97 & 0.577 & 34.83 & 0.245 \\
HFB1 & 0.156 & -15.80 & 244.84 & 0.936 & 28.67 & 0.250 \\
\end{tabular}
\end{ruledtabular}
\end{table}

The SLy4, SLy5, SkM* and SkI3 parameterizations only determine the Skyrme
functional in the particle-hole channel and specify neither the form nor the
parameters of the pairing functional. As mentioned earlier, we use a pairing
functional derived from a density-dependent pairing force with mixed
surface-volume nature, which is controlled by two pairing strengths $V_0^{n}$
and $V_0^{p}$ for neutrons and protons, respectively. Following 
Ref.~\cite{navarroperez2022controlling}, we fit these pairing
strengths to reproduce the 3-point formula for the odd-even mass staggering of
$^{232}$Th.

Since the time-reversal symmetry is broken in the FAM calculation,
we need to consider Skyrme-EDF terms involving time-odd densities: 
\begin{equation}
    \begin{aligned}
        \chi_{t t_{3}}^{\mathrm{(odd)}} =& C_{t}^{s} \vec{s}_{t t_{3}}^{2}
        + C_{t}^{\Delta s} \vec{s}_{t t_{3}} \cdot \nabla^{2} \vec{s}_{t t_{3}} + C_{t}^{j} \vec{j}_{t t_{3}}^{2} \\
        &+C_{t}^{\nabla j} \vec{s}_{t t_{3}} \cdot \nabla \times \vec{j}_{t t_{3}}
        + C_{t}^{T} \vec{s}_{t t_{3}} \cdot \vec{T}_{t t_{3}} \\
        &+ C_{t}^{F} \vec{s}_{t t_{3}} \cdot \vec{F}_{t t_{3}} + C_{t}^{\nabla s}\left(\nabla \cdot \vec{s}_{t t_{3}}\right)^{2}, 
    \end{aligned}
\end{equation}
where we follow the notations given in Ref.~\cite{perlinska2004local}. For 
SLy4, SLy5, SkM* and SkI3, all coupling constants $C_{t}^{u}$ are computed 
from the Skyrme force parameters $(t,x)$ \cite{perlinska2004local}. For HFB1, 
however, only time-even Skyrme couplings are provided in 
\cite{schunck2015error}. We thus first transform these time-even couplings into 
$(t,x)$ parameters and then obtain time-odd couplings from $(t,x)$ 
\cite{schunck2019energy}. Furthermore, in our calculations we set 
$C^{\Delta s}_t=0$ to avoid finite-size instabilities 
\cite{lesinski2006isovector,kortelainen2010instabilities,
schunck2010onequasiparticle,hellemans2013spurious,pastore2015spurious}. 
While this latter choice has very little impact on FAM calculations with SLy4, 
SLy5, SkI3 and SkM*, both in terms of convergence and the values of final 
results, we find that it greatly improves the convergence of calculations 
with HFB1 without significantly changing the results.

Besides the systematic uncertainties, we also study the statistical 
uncertainties of FAM calculations with samples generated from Bayesian model 
calibration. We first calibrate the HFB1 functional following the procedure 
presented in Refs.~\cite{mcdonnell2015uncertainty, higdon2015bayesian, 
schunck2020calibration}, using the the probabilistic programming language Stan \cite{carpenter2017stan}; 
the posterior distribution obtained from the 
calibration is presented in the Supplemental Material \cite{supp}. 
We then propagate the uncertainties of Skyrme 
parameters to FAM responses by carrying out FAM calculations with 50 samples 
taken inside the 95\% credible region of the posterior distribution. 


\section{Multipole Responses in Selected Even-Even Nuclei}
\label{sec:responses_even_even}

\begin{SCfigure*}[0.57][!thb]
\includegraphics[width=0.61\textwidth]{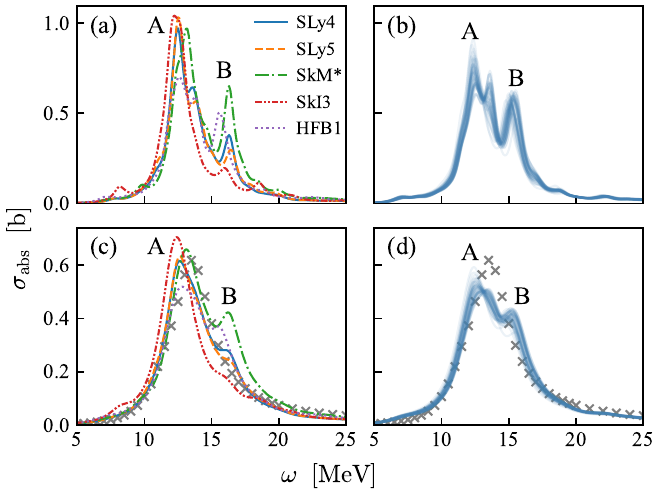}
\caption{Panel (a): photoabsorption cross sections of $^{208}$Pb 
as functions of the photon energy $\omega$, 
calculated with different Skyrme parameterizations and a width of $\Gamma = 1$ MeV. 
Panel (b): similar to panel (a) but calculated with the 50 samples from the HFB1 posterior 
distribution. Panels (c) and (d): similar to panels (a) and (b) but calculated with a 
width of $\Gamma = 2$ MeV. The evaluated nuclear data \cite{goriely2019reference} 
are shown by cross markers. }
\label{fig:Pb208_photo}
\end{SCfigure*}

To benchmark our new code we first perform calculations for the multipole 
responses in doubly magic nucleus $^{208}$Pb, semi-magic spherical nucleus 
$^{90}$Zr, and axially deformed nucleus $^{240}$Pu. Our results are consistent 
with previous FAM calculations \cite{kortelainen2015multipole}, which 
validates the correctness of \textsc{gfam}. In the following we will analyze 
the systematic and statistical uncertainties of the multipole responses in 
these even-even nuclei. 


\subsection{$E1$ transition}
\label{subsec:photo_even_even}

Table \ref{tab:resonance_energies} lists the giant dipole resonance (GDR) 
energies of $^{90}$Zr and $^{208}$Pb obtained from FAM calculations for the 
$E1$ transition. The mean value and standard deviation of the resonance 
energies obtained from calculations with SLy4, SLy5, SkM*, SkI3 and HFB1 are 
taken as the systematic mean, $\bar{x}_{\rm sys}$, and uncertainty, 
$\sigma_{\rm sys}$, respectively, while the statistical mean value 
$\bar{x}_{\rm stat}$ and standard deviation $\sigma_{\rm stat}$ are obtained 
from the posterior samples of Skyrme parameters generated by the Bayesian model 
calibration for HFB1. We see that the calculated GDR energies of $^{90}$Zr and 
$^{208}$Pb are close to their corresponding experimental values. Besides, the 
systematic uncertainty of the GDR energy of $^{90}$Zr is much larger than the 
statistical one, while the two uncertainties for $^{208}$Pb are almost equal. 

\begin{table}[!tbh]
\caption{Giant dipole resonance (GDR) energies of $^{90}$Zr and $^{208}$Pb, and 
giant quadrupole resonance (GQR) energies of $^{208}$Pb obtained from FAM 
calculations with various Skyrme EDFs. For the GQR both isoscalar 
(IS) and isovector (IV) resonance energies are given. The systematic mean value
$\bar{x}_{\rm sys}$ and standard deviation $\sigma_{\rm sys}$ are obtained from 
the results with SLy4, SLy5, SkM*, SkI3 and HFB1 functionals. The statistical 
mean value $\bar{x}_{\rm stat}$ and standard deviation $\sigma_{\rm stat}$ are 
obtained from the 50 HFB1 posterior samples. Experimental 
values for the GDR energies are the averages of the data given in Table 1 of 
\cite{plujko2018giant}; the ISGQR and IVGQR energies are taken from 
\cite{roca-maza2013giant}.}
\label{tab:resonance_energies}
\begin{ruledtabular}
\begin{tabular}{cz{2.1}z{2.1}z{2.1}z{2.1}}
& \multicolumn{1}{c}{$E_{\rm GDR}$($^{90}$\text{Zr})} 
& \multicolumn{1}{c}{$E_{\rm GDR}$($^{208}$\text{Pb})} 
& \multicolumn{1}{c}{$E_{\rm ISGQR}$} 
& \multicolumn{1}{c}{$E_{\rm IVGQR}$} \\
\hline
\text{exp}   & 16.8 & 13.4 & 10.9 & 22.7 \\ 
\hline
\text{SLy4}  & 15.6 & 12.5 & 12.3 & 22.7 \\
\text{SLy5}  & 15.5 & 12.5 & 12.2 & 22.8 \\
\text{SkM*}  & 17.5 & 13.1 & 11.5 & 22.4 \\
\text{SkI3}  & 16.1 & 12.2 & 13.5 & 21.8 \\
\text{HFB1}  & 17.2 & 12.6 & 10.7 & 22.0 \\
\hline
$\bar{x}_{\rm sys}$ & 16.4  & 12.6  & 12.0 & 22.3 \\ 
$\sigma_{\rm sys}$  &  0.8 &  0.3 & 0.9 & 0.4 \\ 
\hline
$\bar{x}_{\rm stat}$ & 17.0  & 12.5  & 12.0  & 21.9 \\
$\sigma_{\rm stat}$  &  0.1 &  0.3 &  0.1 &  0.4 \\
\end{tabular}
\end{ruledtabular}
\end{table}

Figure \ref{fig:Pb208_photo} shows the photoabsorption cross sections of 
$^{208}$Pb calculated with different Skyrme EDFs and widths $\Gamma=1$ 
and 2 MeV. First, we see that the cross sections obtained with different 
parameterizations show similar patterns. The position of the giant resonance 
marked by ``A'' in Fig.~\ref{fig:Pb208_photo} does not vary significantly when 
we change the underlying Skyrme EDF, but its height can differ dramatically, 
especially between HFB1 and other EDFs in panels (a, c) and among the 
statistical samples in panels (b, d). Second, the peak marked by ``B'' in 
Fig.~\ref{fig:Pb208_photo} vary a lot in panels (a, c) as we change the 
underlying EDF; its statistical uncertainty shown in panels (b, d), however, is 
smaller than the systematic one. 

We obtain good agreement with the nuclear data evaluation 
\cite{goriely2019reference} when we take a larger width ($\Gamma=2$ MeV). The 
major difference is that the peak ``B'' (shoulder structure) in 
Fig.~\ref{fig:Pb208_photo} does not exist in the evaluation. This shoulder was 
also observed in previous calculations \cite{nesterenko2007giant,
sasaki2022noniterative} and was attributed to the intruder state with a large 
angular momentum \cite{nesterenko2007giant}. The dependence of peak ``B'' on 
the s.p.\ structure can explain why it is more sensitive to the choice of the 
Skyrme parameterization than the collective resonance ``A''. 

\begin{figure}
\includegraphics[width=0.95\linewidth]{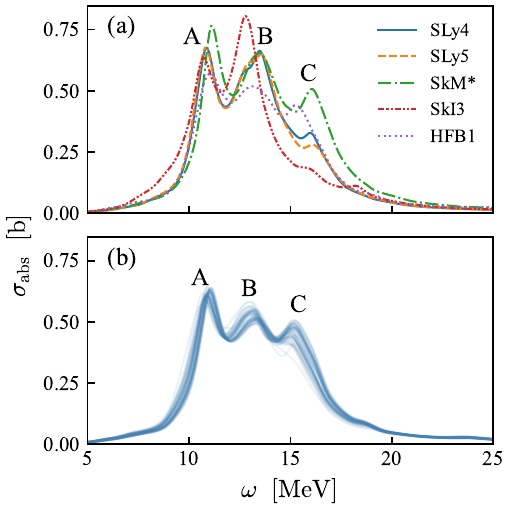}
\caption{Similar to panels (a) and (b) of Fig.~\ref{fig:Pb208_photo} but for 
$^{240}$Pu.}
\label{fig:Pu240_photo}
\end{figure}

The photoabsorption cross section of $^{240}$Pu is displayed in 
Fig.~\ref{fig:Pu240_photo}. As expected, we observe that the ground-state 
deformation of this nucleus causes significant fragmentation of the transition 
strength distribution. The peak ``A'' in Fig.~\ref{fig:Pu240_photo} is given by 
the FAM calculation of $K=0$ and depends weakly on the Skyrme EDF (the only 
outlier is SKM*). The peaks ``B'' and ``C'', however, are quite sensitive to 
the Skyrme parameterizations, and their systematic uncertainties are much 
greater than statistical ones. Both ``B'' and ``C'' peaks originate from the 
strength distribution of $|K|=1$. Since the ground state of $^{240}$Pu has a 
prolate shape, the effective oscillator length along the $z$ axis ($K=0$) is 
longer than that along the $x$ or $y$ axis ($|K|=1$), and thus the peak ``A'' has a 
lower energy than peaks ``B'' and ``C''. 

Interestingly, the variations shown in Figs.~\ref{fig:Pb208_photo} and 
\ref{fig:Pu240_photo} are not strongly correlated with the bulk properties of 
static HFB states: the systematic uncertainty of the HFB energy of $^{240}$Pu 
is 4.6 MeV, which is much smaller than the statistical uncertainty 12.3 MeV. 
However, the photoabsorption cross section shows an opposite trend: 
in Figs.~\ref{fig:Pb208_photo} and \ref{fig:Pu240_photo} the systematic 
uncertainty is much larger than the statistical one. 
This offers conclusive 
evidence that observables related to nuclear electromagnetic transitions can 
provide information that is not captured by the fits of EDFs with ground-state 
properties only.


\subsection{Sum rules}
\label{subsec:sum_rule}

To better understand the variations of the photoabsorption cross sections 
discussed in the previous section, we study the systematic and statistical 
uncertainties of sum rules \cite{bohigas1979sum,lipparini1989sum,ring2004nuclear} 
\begin{equation}
    m_k(F) = \int \omega^k \frac{dB(\omega; F)}{d\omega} d\omega 
    = \sum_{n>0} \Omega_n^k \left| \left\langle n|F|0 \right\rangle \right|^2. 
\end{equation}

We first look into the energy-weighted sum rule $m_1$, which can be evaluated 
with the Hamiltonian $H$ and static HFB state $|0\rangle$ as 
\cite{thouless1961vibrational,bertsch1975study,bohigas1979sum,lipparini1989sum,
ring2004nuclear,khan2002continuum,hinohara2015complexenergy}
\begin{equation}\label{eq:m1_def}
    m_1(F) = \frac{1}{2} \langle 0| [F^\dagger, [H, F]] |0 \rangle
    = (1+\kappa) m_1^T(F), 
\end{equation}
where $m_1^T$ is given by the double commutator involving the kinetic-energy 
term only, and $\kappa$ is the enhancement factor due to the momentum dependence 
of the effective interaction. For the Skyrme EDF, the
expressions of $m_1$ for electric multipole operators are presented in Appendix 
\ref{app:sum_rule}, based on which we can obtain the TRK sum rule 
for the $E1$ transition as \cite{lipparini1989sum,ring2004nuclear,oishi2016finite}
\begin{equation}\label{eq:m1_E1}
    m_1(E1) = \frac{e^2 \hbar^2}{2m} \left[ 1+\kappa(E1) \right] \frac{NZ}{A} \frac{3}{4\pi},
\end{equation}
where the enhancement factor $\kappa(E1)$ is the only component that depends on 
the static HFB solution 
\begin{equation}\label{eq:kappa_E1}
    \kappa(E1) = \frac{2m}{\hbar^2}\frac{A}{NZ} \left( C^\tau_0 - C^\tau_1 \right)
    \int \rho_n(\vec{r}) \rho_p(\vec{r}) d\vec{r},  
\end{equation}
where $\rho_n$ and $\rho_p$ are neutron and proton densities of the HFB ground 
state, and $C^\tau_t$ is the coupling constant of the term $\rho_t \tau_t$ in the 
Skyrme EDF. Based on Eq.~\eqref{eq:kappa_E1}, we can write the TRK enhancement factor 
in symmetric nuclear matter as \cite{klupfel2009variations,oishi2016finite}
\begin{equation}\label{eq:kappa_NM}
    \kappa_{\rm TRK} = \frac{2m}{\hbar^2}\left( C^\tau_0 - C^\tau_1 \right) \rho_c 
    = M_v^{*-1} - 1, 
\end{equation}
where $\rho_c$ is the saturation density and $M_v^*$ is the isovector effective 
mass in the unit of the nucleon mass $m$. The ratio of enhancement factors 
\eqref{eq:kappa_E1} and \eqref{eq:kappa_NM} is then
\begin{equation}\label{eq:kappa_ratio}
    \frac{\kappa(E1)}{\kappa_{\rm TRK}} = \frac{A}{NZ \rho_c} \int \rho_n(\vec{r}) \rho_p(\vec{r}) d\vec{r}. 
\end{equation}

Table \ref{tab:m1_Pb208} lists the energy-weighted sum rules and enhancement 
factors for the $E1$ transition in $^{208}$Pb, obtained from static HFB solutions  
calculated with various Skyrme EDFs. Integrating energy-weighted 
FAM strength distributions up to 50 MeV can give 96\% to 98\% of 
the sum rules presented in Table \ref{tab:m1_Pb208}. 

\begin{table}[tbh]
\caption{Energy-weighted sum rules $m_1(E1)$ \eqref{eq:m1_def} in $e^2\mathrm{b}\:\mathrm{MeV}$ 
for $^{208}$Pb, as well as enhancement factors $\kappa(E1)$ \eqref{eq:kappa_E1}, $\kappa_{\rm TRK}$ 
\eqref{eq:kappa_NM}, and their ratios \eqref{eq:kappa_ratio}. 
They are obtained from the HFB ground states calculated with various 
Skyrme EDFs. See the caption of Table \ref{tab:resonance_energies} for the 
definitions of $\bar{x}$ and $\sigma$. }
\label{tab:m1_Pb208}
\begin{ruledtabular}
\begin{tabular}{cz{1.3}z{1.3}z{1.3}z{1.3}}
& \multicolumn{1}{c}{$m_1(E1)$} 
& \multicolumn{1}{c}{$\kappa(E1)$} 
& \multicolumn{1}{c}{$\kappa_{\rm TRK}$}
& \multicolumn{1}{c}{$\kappa(E1)/\kappa_{\rm TRK}$} \\
\hline
SLy4  & 2.909 & 0.189 & 0.250 & 0.750 \\
SLy5  & 2.910 & 0.189 & 0.250 & 0.760 \\
SkM*  & 3.428 & 0.401 & 0.532 & 0.750 \\
SkI3  & 2.920 & 0.187 & 0.245 & 0.760 \\
HFB1  & 2.933 & 0.193 & 0.250 & 0.770 \\
\hline
$\bar{x}_{\rm sys}$ & 3.020  & 0.232  & 0.305 & 0.758 \\ 
$\sigma_{\rm sys}$ & 0.204 & 0.085 & 0.113 & 0.007 \\ 
\hline
$\bar{x}_{\rm stat}$ & 2.935 & 0.194 & 0.250 & 0.780 \\
$\sigma_{\rm stat}$ & 0.005 & 0.002 & 0.000 &  0.008 \\
\end{tabular}
\end{ruledtabular}
\end{table}

In Table \ref{tab:m1_Pb208} we first note that the ratio \eqref{eq:kappa_ratio} 
is nearly constant, which indicates that ground-state density profiles have 
limited variations \cite{oishi2016finite}. Therefore, the energy-weighted sum 
rule for a given nucleus is almost solely determined by the isovector effective 
mass. In the Bayesian calibration of HFB1, the isovector effective mass is 
fixed at the same value as SLy4 and SLy5 \cite{kortelainen2010nuclear,
kortelainen2012nuclear,schunck2015error}: this explains why the statistical 
uncertainties of $m_1(E1)$ and $\kappa(E1)$ are very small. We find that the 
values of $m_1(E1)$ are very similar for nearly all the EDFs. This is 
the reason why the differences in photoabsorption cross sections in 
Fig.~\ref{fig:Pb208_photo} are rather small, since their variations must be 
constrained by the sum rule $m_1(E1)$. The only exception is SkM*, which has a substantially 
larger value of $m_1(E1)$ that manifests itself by a strong peak ``B'' in 
Fig.~\ref{fig:Pb208_photo}.

Besides $m_1$, another interesting quantity worth investigating is the average peak energy 
defined in two different ways \cite{bohigas1979sum}: 
\begin{equation}\label{eq:avg_E1_peak_pos}
    \overline{\omega}(F) = \frac{m_1(F)}{m_0(F)} \geq 
    \widetilde{\omega}(F) = \sqrt{\frac{m_1(F)}{m_{-1}(F)}}. 
\end{equation}
Here we focus on the first average energy $\overline{\omega}$,
while discussions on $\widetilde{\omega}$ can be found in the Supplemental Material \cite{supp}. 
In Ref.~\cite{nesterenko2007giant}, calculations with a separable RPA approach and 
a few Skyrme parameterizations suggested that $\overline{\omega}$ for the $E1$ 
transition is strongly correlated with 
\begin{equation}\label{eq:frequency_NM_surf_isospin}
    \overline{\Omega}^{\rm NM}_{\rm surf} 
    = \sqrt{\left( a_{\rm sym}^{\rm NM} - \frac{L_{\rm sym}^{\rm NM}}{6} \right)
    M_v^{*-1}}, 
\end{equation}
where $a_{\rm sym}^{\rm NM}$ is the symmetry energy at saturation 
density, and $L_{\rm sym}^{\rm NM}$ is the density derivative of 
$a_{\rm sym}^{\rm NM}$ (see Table \ref{tab:nm} for their values). 
The quantity $\left( a_{\rm sym}^{\rm NM} - L_{\rm sym}^{\rm NM}/6 \right)$ 
is approximately the symmetry energy at the nuclear surface with density 
$\rho=\rho_c/2$. Thus, 
$\overline{\Omega}^{\rm NM}_{\rm surf}$ can be understood as the oscillating 
frequency of the isovector density $\rho_1$ around zero at the nuclear surface. 
Since the $E1$ operator is mostly isovector, it is natural that the frequency 
$\overline{\Omega}^{\rm NM}_{\rm surf}$ is highly correlated with the average $E1$ 
peak position. 

\begin{figure}
\includegraphics[width=0.95\linewidth]{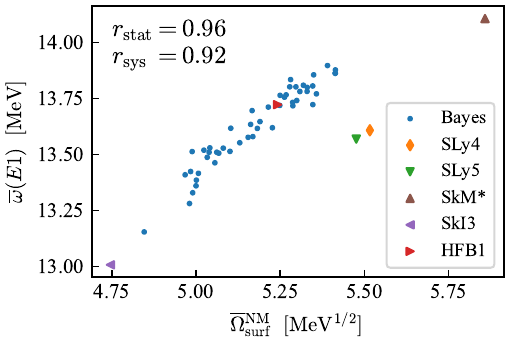}
\caption{Correlation between the nuclear-surface isovector oscillating frequency 
$\overline{\Omega}^{\rm NM}_{\rm surf}$ \eqref{eq:frequency_NM_surf_isospin} 
and the average $E1$ peak position $\overline{\omega}(E1)$ \eqref{eq:avg_E1_peak_pos} in $^{208}$Pb, 
obtained from FAM calculations with different Skyrme EDFs. The correlation 
coefficient (CC) of the two quantities calculated with the statistical Bayesian 
(Bayes) samples from the HFB1 posterior distribution is $r_{\rm stat}=0.96$, 
while the CC across the five EDFs that are used for the analysis of 
systematic uncertainties is $r_{\rm sys}=0.92$. }
\label{fig:Pb208_avg_omega}
\end{figure}

Figure \ref{fig:Pb208_avg_omega} plots the value of $\overline{\omega}(E1)$ as 
a function of $\overline{\Omega}^{\rm NM}_{\rm surf}$ in $^{208}$Pb. We observe 
a very strong correlation between these two quantities for the statistical 
samples from the HFB1 posterior distribution, with a Pearson correlation coefficient of $r_{\rm stat}=0.96$. 
However, the correlation across functionals SLy4, SLy5, SkM*, SkI3 and HFB1 
in Fig.~\ref{fig:Pb208_avg_omega} is weaker ($r_{\rm sys} = 0.92$), 
with the points of SLy4, SLy5, and SkM* lying away from the line formed by the statistical samples. 
This weaker correlation suggests that other effects such as the isoscalar components of the EDF have an impact. 
The correlation shown in Fig.~\ref{fig:Pb208_avg_omega} demonstrates that information on $E1$ transitions can help 
us constrain the asymmetric (isovector) nuclear-matter properties in the Skyrme 
EDF. The sum rules given by FAM calculations for the (deformed) $^{240}$Pu nucleus give 
a similar correlation plot as Fig.~\ref{fig:Pb208_avg_omega}. 


\subsection{High-order multipole responses}
\label{subsec:high_order_Pb_Pu}

Table \ref{tab:resonance_energies} shows that the isoscalar and isovector giant 
quadrupole resonance (GQR) energies of $^{208}$Pb are close to the experimental 
values. In addition, the systematic uncertainty of the isoscalar GQR energy is much 
larger than the statistical one, while systematic and statistical uncertainties 
of the isovector GQR location are similar. 

\begin{figure}[t!htb]
\includegraphics[width=0.95\linewidth]{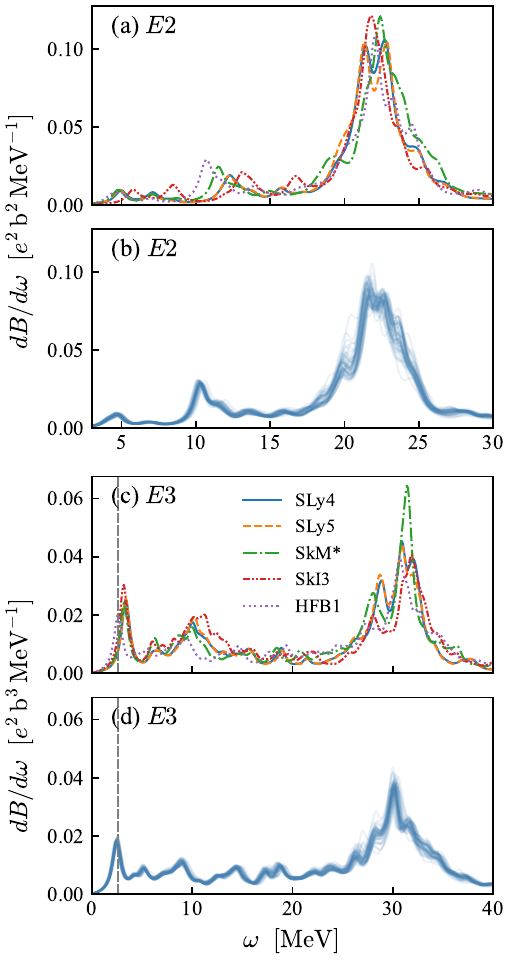}
\caption{Panel (a): isovector electric quadrupole ($E2$) responses in $^{208}$Pb
calculated with different Skyrme EDFs. Panel (b): similar to panel (a) but 
calculated with the 50 samples from the HFB1 posterior distribution. Panels (c) and (d): 
similar to panels (a) and (b) but for isovector electric octupole ($E3$) responses. 
The excitation energy of the first $3^-$ state of $^{208}$Pb \cite{martin2007nuclear} 
is marked by a dashed vertical line in panels (c) and (d). A width of $\Gamma=1$ MeV
is adopted in all the panels. }
\label{fig:Pb208_E2E3}
\end{figure}

Figure \ref{fig:Pb208_E2E3} shows the strength distributions of isovector 
$E2$ and $E3$ transitions in $^{208}$Pb; see the Supplemental Material \cite{supp} 
for isoscalar $E2$ and $E3$ responses in $^{208}$Pb. These 
high-order responses show features similar to the $E1$ response, with the 
systematic uncertainty significantly larger than the statistical one. However, 
we also observe multiple small peaks below the giant resonance. Some of these 
peaks are sensitive to the EDF choice, since they are less collective than the 
giant resonance and depend heavily on the s.p.\ structure. It is worth mentioning that the 
position of the lowest peak in the isovector $E3$ response is close to the first $3^-$ 
state of $^{208}$Pb \cite{martin2007nuclear}, which validates the 
effectiveness of our calculations. 

To better understand Fig.~\ref{fig:Pb208_E2E3}, we report the energy-weighted 
sum rules for $E2$ and $E3$ transitions in Table \ref{tab:m1_E2E3M1_Pb208}. 
Integrating energy-weighted $E2$ and $E3$ strength distributions up to 50 MeV 
can give more than 95\% of the sum rules listed in Table \ref{tab:m1_E2E3M1_Pb208}. 
Similar to the $E1$ transition, the values of $m_1(E2)$ and $m_1(E3)$ calculated with 
different Skyrme EDFs are close to each other, with SkM* being the only outlier 
due to its large enhancement factor $\kappa_{\rm TRK}$. This explains why the 
strength distributions of SkM* show higher and stronger $E2$ and $E3$ 
resonances. 

\begin{table}[!tbh]
\caption{Energy-weighted sum rules $m_1$ for $E2$, $E3$, and $M1$ transitions 
in $^{208}$Pb, in the units of $e^2\mathrm{b}^2\mathrm{MeV}$, 
$e^2\mathrm{b}^3\mathrm{MeV}$, and $\mu_{\rm N}^2\mathrm{MeV}$, 
respectively. All the values for electric transitions are obtained from the HFB 
ground states calculated with various Skyrme EDFs. For the $M1$ transition, sum 
rules evaluated via Eq.~\eqref{eq:m1_M1_Eso} and via integrating the 
energy-weighted FAM strength distribution are both presented. 
See the caption of Table \ref{tab:resonance_energies} for the 
definitions of $\bar{x}$ and $\sigma$. }
\label{tab:m1_E2E3M1_Pb208}
\begin{ruledtabular}
\begin{tabular}{cz{2.2}z{2.2}z{3.2}z{3.2}}
& \multicolumn{1}{c}{$m_1(E2)$} 
& \multicolumn{1}{c}{$m_1(E3)$} 
& \multicolumn{1}{c}{$m_1^{\rm SO}(M1)$}
& \multicolumn{1}{c}{$m_1(M1)$} \\
\hline
SLy4 & 12.10 &  9.78 & 509.90 & 515.10 \\
SLy5 & 12.06 &  9.72 & 519.30 & 617.50 \\
SkI3 & 12.20 &  9.90 & 509.10 & 508.80 \\
SkM* & 13.84 & 10.92 & 556.50 & 563.70 \\
HFB1 & 12.24 &  9.88 & 424.30 & 434.60 \\
\hline
$\bar{x}_{\rm sys}$ & 12.49 & 10.04 & 503.82 & 527.94 \\
$\sigma_{\rm sys}$  &  0.68 &  0.44 &  43.37 &  60.90 \\
\hline
$\bar{x}_{\rm stat}$ & 12.24 & 9.88 & 432.60 & 443.90 \\
$\sigma_{\rm stat}$  &  0.03 & 0.06 &  16.00 &  15.90 \\
\end{tabular}
\end{ruledtabular}
\end{table}

Now we turn to the analysis of magnetic transitions. The strength distributions 
of isovector $M1$ and $M2$ transitions in $^{208}$Pb are shown in 
Fig.~\ref{fig:Pb208_M1M2}; see the Supplemental Material \cite{supp} for 
isoscalar magnetic responses in $^{208}$Pb. We notice that the systematic 
uncertainties of magnetic transitions are much larger than those of electric 
transitions while the statistical uncertainties remain small. This observation 
is consistent with the results of Ref.~\cite{sasaki2022noniterative}, and 
agrees with the argument that predicting $M1$ giant resonances is challenging 
for Skyrme EDFs \cite{nesterenko2010spinflip}. 

\begin{figure}[!htb]
    \includegraphics[width=0.95\linewidth]{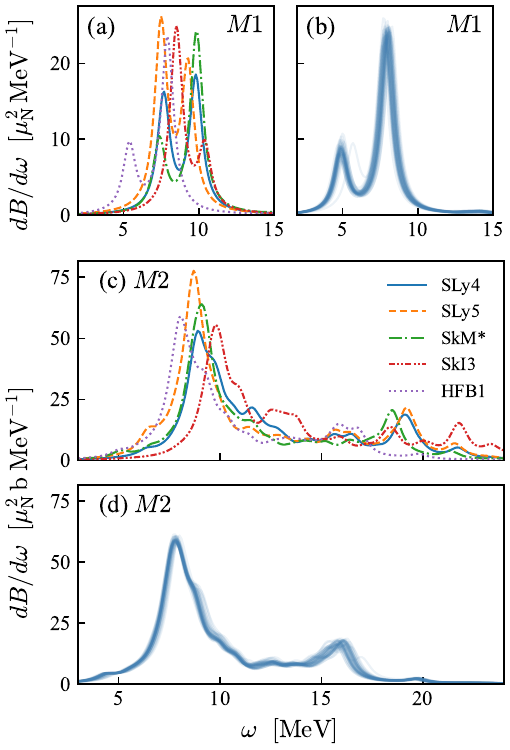}
    \caption{Similar to Fig.~\ref{fig:Pb208_E2E3} but for isovector 
    magnetic dipole ($M1$) and quadrupole ($M2$) responses in $^{208}$Pb. }
    \label{fig:Pb208_M1M2}
\end{figure}

Table \ref{tab:m1_E2E3M1_Pb208} also presents the energy-weighted sum rules 
evaluated with the FAM response function for the $M1$ transition. We can see 
that $m_1(M1)$ has a larger systematic uncertainty than the statistical one, 
which is consistent with Fig.~\ref{fig:Pb208_M1M2}. In principle we can 
calculate $m_1(M1)$ via Eq.~\eqref{eq:m1_def}, but the corresponding expression 
is too complicated for our analysis \cite{traini1978energyweighted}. 
Instead, we use the HFB spin-orbit energy $E_{\rm SO}$ to calculate the 
Kurath sum rule \cite{kurath1963strong,lipparini1989sum}
\begin{equation}\label{eq:m1_M1_Eso}
    m_1^{\rm SO}(M1) = -\frac{3}{16\pi}\left( g_s^{(n)} + g_s^{(p)} \right)^2 \mu_{\rm N}^2 E_{\rm SO}, 
\end{equation}
which provides the largest contribution to the exact sum rule $m_1(M1)$, 
especially when the tensor term is not present. The quantity $m_1^{\rm SO}(M1)$ 
shown in Table \ref{tab:m1_E2E3M1_Pb208} has again a large systematic but a 
small statistical uncertainty, in agreement with the the exact sum rule 
$m_1(M1)$. Therefore, the systematic uncertainty for the $M1$ response can be 
partly attributed to the uncertainty of the HFB spin-orbit energy, which is 
expected to be sensitive to the s.p.\ structure. In contrast, the electric 
multipole resonances are more collective and less dependent on the details 
of s.p.\ levels. Moreover, we notice in Table \ref{tab:m1_E2E3M1_Pb208} that 
the inclusion of the tensor term in SLy5 significantly enlarges the difference 
between the Kurath and exact sum rules, making Eq.~\eqref{eq:m1_M1_Eso} a worse 
approximation for $m_1(M1)$ when the tensor term is present. 


\section{Electromagnetic Responses in Actinide Nuclei}
\label{sec:actinides}

In this section, we review the electromagnetic response properties -- either 
the photoabsorption cross sections or high-order responses -- in actinide 
nuclei. These include even-even, odd-mass and odd-odd nuclei.


\subsection{Odd-mass and odd-odd isotopes}
\label{subsec:blocking}

In odd-$A$ and odd-odd nuclei, the HFB solution on top of which the nuclear 
response is computed depends on the blocking configuration(s), which sets the 
spin and parity of the nucleus. For most EDFs, it is not guaranteed that the 
lowest-energy blocking configuration gives the observed spin and parity of the 
nucleus. In this section, we use $^{239}$U (odd-$A$) and $^{238}$Np (odd-odd) 
to study the sensitivity of multipole responses to the choice of underlying 
blocking configurations. To eliminate uncertainties related to the parameters 
of the EDF, we only consider the SLy4 parameterization in the following.


\subsubsection{Ground-state energies and sum rules}
\label{subsubsec:global}

In \textsc{hfbtho} the dimensionless quadrupole deformation $\beta_2$ is 
defined as $\beta_2 = \sqrt{\pi/5} \langle 2z^2-x^2-y^2\rangle / \langle r^2 \rangle$, 
where $\langle \cdots \rangle$ represents the 
expectation value in the static HFB state. Table \ref{tab:U239_blocking} gives 
the HFB energy and quadrupole deformation $\beta_2$ of $^{239}$U for different 
neutron blocked orbits. While the impact of the blocking configuration on the 
ground-state energy is of the order of a few hundreds of keV, the deformation 
of $^{239}$U does not vary much, except when we block $[770]1/2^-$, an intruder 
orbit that has a large value of $n_z$ and thus a density profile concentrated 
along the $z$ axis. The same observation can also be made for different 
combinations of neutron and proton blocked orbits in $^{238}$Np, where 
blocking the intruder neutron state $[770]1/2^-$ leads to a larger HFB 
deformation. 

\begin{table}[!htb]
\caption{HFB energies $E_{\rm HFB}$ in MeV, quadrupole deformations $\beta_2$, 
and energy-weighted sum rules $m_1$ for $E1$ and $E2$ ($K=0, 2$) transitions
in the units of $e^2\mathrm{b}\:\mathrm{MeV}$ and $e^2\mathrm{b}^2\mathrm{MeV}$, 
obtained with various neutron blocked orbits in $^{239}$U. We use the SLy4 
parameterization and $^{238}$U as the core for blocking. Blocked orbits  
are labeled by their asymptotic (Nilsson) quantum numbers 
$[N n_z \Lambda] \Omega^\pi$. The uncertainty $\sigma_{\rm blck.}$ is 
given by the standard deviation of the results obtained with different blocked 
orbits. }
\label{tab:U239_blocking}
\begin{ruledtabular}
\begin{tabular}{cz{5.2}z{1.3}z{1.4}z{2.2}z{2.2}}
& \multicolumn{1}{c}{$E_{\rm HFB}$} 
& \multicolumn{1}{c}{$\beta_2$} 
& \multicolumn{1}{c}{$m_1(E1)$} 
& \multicolumn{2}{c}{$m_1(E2)$} \\
& & & & \multicolumn{1}{c}{$K=0$} & \multicolumn{1}{c}{$K=2$} \\
\hline
$[6 2 2]5/2^+$ & \text{-}1794.72 & 0.273 & 3.3124 & 18.61 & 13.06 \\
$[7 4 3]7/2^-$ & \text{-}1794.67 & 0.268 & 3.3122 & 18.53 & 13.10 \\
$[6 2 4]7/2^-$ & \text{-}1794.37 & 0.265 & 3.3118 & 18.50 & 13.13 \\
$[6 3 1]1/2^+$ & \text{-}1793.95 & 0.263 & 3.3122 & 18.47 & 13.14 \\
$[7 7 0]1/2^-$ & \text{-}1793.76 & 0.290 & 3.3119 & 18.86 & 12.95 \\
$\sigma_{\rm blck.}$     &  0.38    & 0.010 & 0.0002 & 0.14  & 0.07 \\   
\end{tabular}
\end{ruledtabular}
\end{table}


\subsubsection{$E1$ transition}
\label{subsubsec:blocking_E1}

\begin{figure}
    \includegraphics[width=0.95\linewidth]{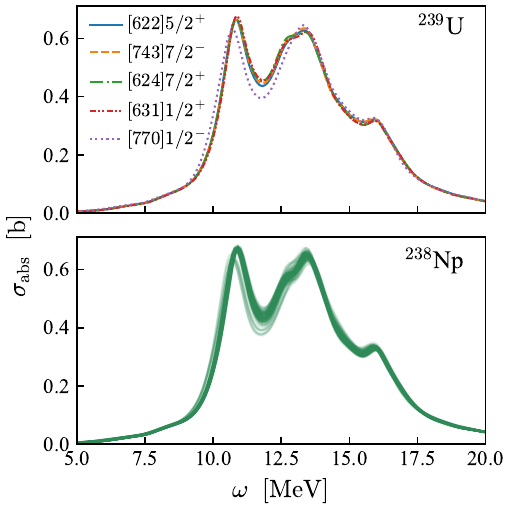}
    \caption{Photoabsorption cross sections of $^{239}$U (upper panel) and 
    $^{238}$Np (lower panel), calculated with the SLy4 parameterization and 
    different blocked orbits. Neutron blocked states in $^{239}$U are 
    labeled by asymptotic quantum numbers $[N n_z \Lambda] \Omega^\pi$. 
    Neutron and proton blocked orbits in $^{238}$Np are given in the Supplemental Material \cite{supp}.
    A width of $\Gamma=1$ MeV is adopted in both panels. }
    \label{fig:U239_Np238_E1}
\end{figure}

Figure \ref{fig:U239_Np238_E1} presents the $E1$ photoabsorption cross sections 
of $^{239}$U and $^{238}$Np obtained with various blocked orbits. First, we see 
that the uncertainty brought by the choice of blocked orbits is much smaller 
than the systematic uncertainty shown in Sec.~\ref{subsec:photo_even_even}. 
In $^{239}$U (upper panel of Fig.~\ref{fig:U239_Np238_E1}), the outlier among 
all the curves is given by the intruder state $[770]1/2^-$: the peak at 
$\omega = 10 \sim 11$ MeV, which originates from the $K=0$ component, slightly 
moves leftward when we block $[770]1/2^-$. This shift can be attributed to the 
larger HFB deformation brought by the intruder state, which results in a longer 
effective oscillator length in the $z$ direction. On the other hand, the 
$|K|=1$ component that gives peaks at higher energies is less impacted by the 
choice of the blocked orbit, indicating that the effective oscillator lengths 
along $x$ and $y$ axes stay stable no matter which state we block. The same 
conclusion can also be drawn from the cross sections of $^{238}$Np. 

The energy-weighted sum rules $m_1$ for the $E1$ transitions in $^{239}$U 
calculated with different blocked states are also given in Table 
\ref{tab:U239_blocking}. We notice that the sum rule $m_1(E1)$ has very weak 
dependence on the choice of the blocked orbit, indicating that the 
corresponding HFB solutions share similar density profiles that enter 
Eq.~\eqref{eq:kappa_E1}. This places a constraint on how much variation the 
cross section can have in Fig.~\ref{fig:U239_Np238_E1}. 


\subsubsection{$E2$ and $M1$ transitions}
\label{subsubsec:blocking_E2_M1}

\begin{figure}[!htb]
    \includegraphics[width=0.95\linewidth]{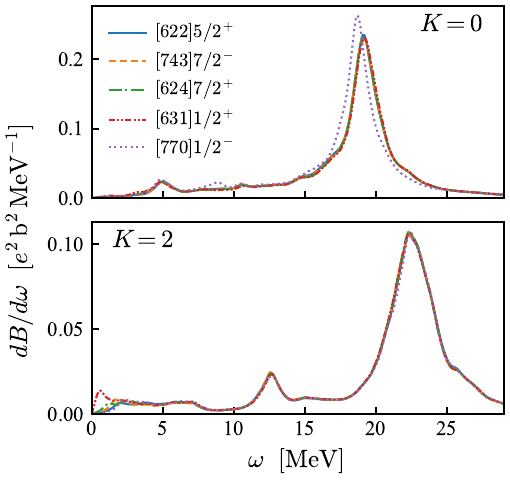}
    \caption{Similar to Fig.~\ref{fig:U239_Np238_E1}, 
    but for $K=0$ and $K=2$ contributions to the isovector $E2$ response in $^{239}$U. }
    \label{fig:U239_E2}
\end{figure}

\begin{figure}[!htb]
    \includegraphics[width=0.95\linewidth]{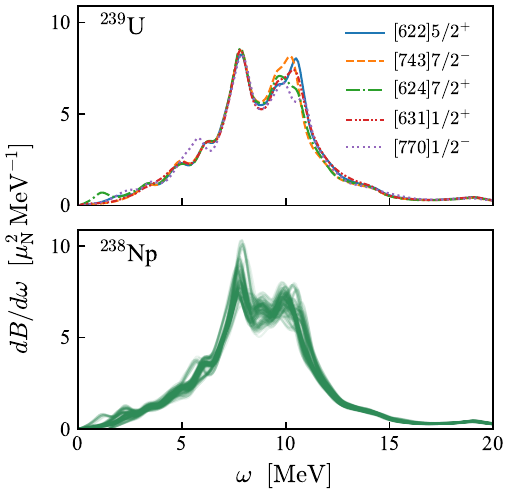}
    \caption{Similar to Fig.~\ref{fig:U239_Np238_E1}  
    but for the $K=0$ components of isovector $M1$ responses. }
    \label{fig:U239_Np238_M1}
\end{figure}

\begin{figure*}
\includegraphics[width=0.95\linewidth]{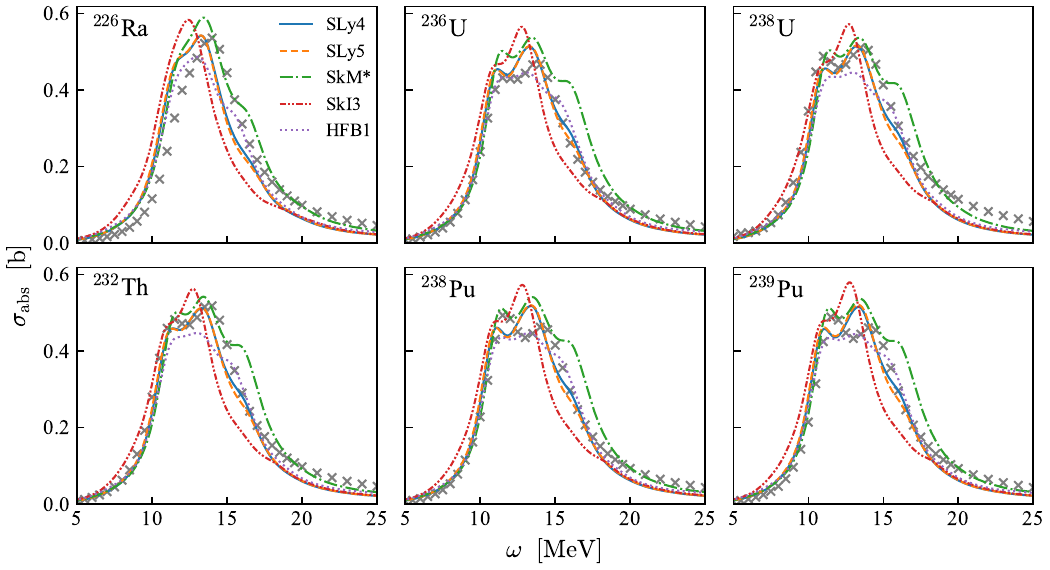}
\caption{Photoabsorption cross sections of $^{226}$Ra, $^{232}$Th, 
$^{236,238}$U and $^{238,239}$Pu, calculated with different Skyrme parameterizations
and a width of $\Gamma=2$ MeV. The evaluated nuclear data 
\cite{goriely2019reference} are denoted by cross markers.}
\label{fig:actinides_exp}
\end{figure*}

Figure \ref{fig:U239_E2} shows the dependency of the isovector $E2$ response on the 
blocking configuration. For the $K=0$ component, the $E2$ resonance is shifted
to a lower energy when the intruder state $[770]1/2^-$ is blocked, while other 
blocking configurations give nearly identical responses. When the 
angular-momentum projection $K$ increases, the dependence on the blocking 
configuration becomes even weaker. The $K=2$ component shown in the lower 
panel of Fig.~\ref{fig:U239_E2} hardly changes when we block
different orbits, except at $\omega$ around 1 MeV. These observations 
also apply to the $E2$ response in $^{238}$Np, and they are consistent with 
what we see in Sec.~\ref{subsubsec:blocking_E1} for $E1$ transitions. 

Unlike the $E1$ transition, Table \ref{tab:U239_blocking} shows that the 
energy-weighted sum rules for the $E2$ transition in $^{239}$U have some dependence 
on the blocked orbit. The sum rule for the $K=0$ component varies significantly 
when the $[770]1/2^-$ orbit is blocked, since 
Eq.~\eqref{eq:nabla_f_20_sqr} shows that it is sensitive to how far the nuclear 
density extends in the $z$ direction. This explains the stronger resonance of 
$[770]1/2^-$ in Fig.~\ref{fig:U239_E2}. The sum rule for the $K=2$ component, 
however, is less sensitive to blocking; according to 
Eq.~\eqref{eq:nabla_f_22_sqr}, it depends on how extended the nucleus is in  
$x$ and $y$ directions, which barely changes when we block different states. 

Figure \ref{fig:U239_Np238_M1} displays the dependence of $M1$ responses 
($K=0$) on the blocked states in $^{239}$U and $^{238}$Np. Overall, the magnetic 
response is much more sensitive to the blocking configuration. Together, 
Figs.~\ref{fig:Pb208_M1M2} and \ref{fig:U239_Np238_M1} suggest strong 
dependence of $M1$ transitions on the s.p.\ structure. Therefore, magnetic 
responses have large uncertainties related to the EDF parameterization and the choice of 
blocked orbits. 


\subsection{Photoabsorption cross sections of major actinides}
\label{subsec:actinides}

In this section we study the photoabsorption cross sections of major actinides 
and their uncertainties. With the help of monoenergetic, high-intensity $\gamma$-ray 
sources such as TUNL \cite{weller2009research}, fission properties of actinides 
can be measured with very high accuracy as functions of excitation energy in 
photon-induced reactions; see, e.g., \cite{krishichayan2018monoenergetic,
krishichayan2019fission,finch2021measurements} for recent examples. At the same 
time, radiative processes such as $(n,\gamma)$ that are essential in nuclear 
astrophysical simulations can be probed indirectly from photonuclear reactions 
$(\gamma,n)$. Such calculations begin with the determination of the total 
photoabsorption cross section. 

Since the total uncertainty is dominated by the systematic one, as shown in 
Secs.~\ref{sec:responses_even_even} and \ref{subsec:blocking}, we only perform 
calculations for the five different functionals used throughout this paper. 
Figure \ref{fig:actinides_exp} shows the photoabsorption cross sections of 
$^{226}$Ra, $^{232}$Th, $^{236,238}$U and $^{238,239}$Pu. Like Fig.~\ref{fig:Pb208_photo} for $^{208}$Pb, 
FAM calculations for nuclei in Fig.~\ref{fig:actinides_exp} are performed with 
$\Gamma=2$ MeV for comparison with the reference database 
\cite{goriely2019reference}. Note that $^{226}$Ra has an octupole-deformed (parity-breaking) 
ground state \cite{cao2020landscape} and $^{239}$Pu is an odd-mass isotope with spin 1/2$^{+}$. Table 
\ref{tab:Pu239_blocking} shows the quantum numbers of blocking configurations 
for the ground state of $^{239}$Pu calculated with different Skyrme parameterizations: 
only with SkM* do we obtain the proper spin-parity assignment of the ground state, yet other 
functionals such as SLy4, SLy5 and HFB1 better reproduce the cross section data.

\begin{table}[!tbh]
\caption{Ground-state blocked orbits in $^{239}$Pu for various Skyrme EDFs, 
labeled by asymptotic (Nilsson) quantum numbers $[N n_z \Lambda] \Omega^\pi$. 
The even-even nucleus core is $^{240}$Pu. }
\label{tab:Pu239_blocking}
\begin{ruledtabular}
\begin{tabular}{ccccc}
\text{SLy4} & \text{SLy5} & \text{SkM*} & \text{SkI3} & \text{HFB1} \\
    \hline
\text{[7,4,3]}$\tfrac{7}{2}^-$ & 
\text{[6,2,2]}$\tfrac{5}{2}^+$ & 
\text{[6,3,1]}$\tfrac{1}{2}^+$ & 
\text{[6,2,4]}$\tfrac{7}{2}^+$ & 
\text{[7,4,3]}$\tfrac{7}{2}^-$ \\
\end{tabular}
\end{ruledtabular}
\end{table}

Overall, we obtain good agreement between FAM predictions and reference 
evaluations, with the results of SkM* and SkI3 being slightly worse than other 
EDFs. Furthermore, the cross sections of different nuclei share similar 
patterns; the similarity is most pronounced when we compare the results of two 
U (or Pu) isotopes.

\begin{figure}[!htb]
\includegraphics[width=0.95\linewidth]{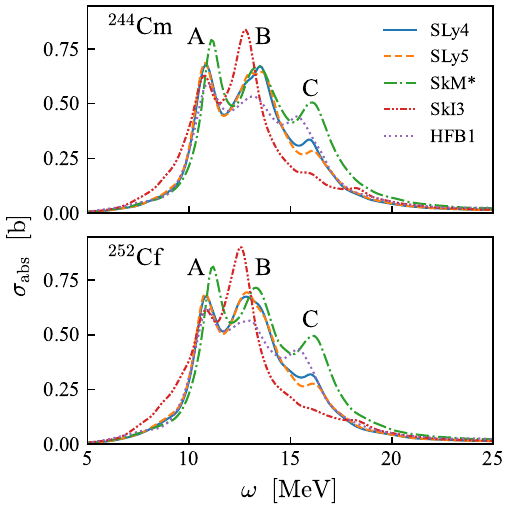}
\caption{Photoabsorption cross sections of $^{244}$Cm and $^{252}$Cm  
calculated with different Skyrme EDFs and a width of $\Gamma=1$ MeV. }
\label{fig:actinides}
\end{figure}

\begin{figure}[!htb]
\includegraphics[width=0.95\linewidth]{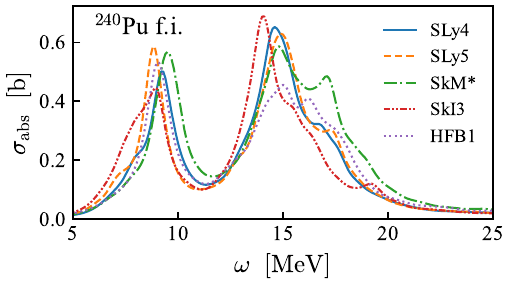}
\caption{Similar to Fig.~\ref{fig:actinides} but for the 
fission isomer (f.i.) of $^{240}$Pu. }
\label{fig:Pu240fi}
\end{figure}

In Figs.~\ref{fig:actinides} and \ref{fig:Pu240fi}, 
we show the photoabsorption cross sections of 
$^{244}$Cm, $^{252}$Cf, and the fission isomer (f.i.) of $^{240}$Pu. For these 
calculations, we adopt $\Gamma = 1$ MeV to better highlight individual peaks. 
We note that the cross sections of $^{244}$Cm and $^{252}$Cf have very 
similar patterns. In Fig.~\ref{fig:actinides} peaks ``B'' and ``C'' ($|K|=1$) have larger 
systematic uncertainties than the peak ``A'' ($K=0$), which is consistent with 
Fig.~\ref{fig:Pu240_photo} for $^{240}$Pu. By comparing Figs.~\ref{fig:Pu240_photo} 
and \ref{fig:Pu240fi}, we see that the photoabsorption cross section of the 
$^{240}$Pu fission isomer is significantly different from that of the ground state. 
This is largely determined by the fact that the fission isomer has a much larger deformation 
($\beta_2 \approx 0.7$) than the ground state, which leads to stronger fragmentation 
in Fig.~\ref{fig:Pu240fi}. 


\subsection{Multipole responses along the plutonium isotopic chain}
\label{subsec:gsf_pu}

The \textsc{gfam} code is an efficient tool for large-scale studies of 
multipole responses through the nuclear landscape. As an example of its 
capability, in this section we present the FAM responses in even-even 
plutonium isotopes from the two-proton to the two-neutron dripline. In this 
section, all HFB and FAM calculations were performed with the Skyrme parameterization SLy4. 
In this case, we find that the two-proton dripline is located at 
$N=122$ and the two-neutron dripline at $N=210$.

\begin{figure}[!htb]
    \includegraphics[width=0.95\linewidth]{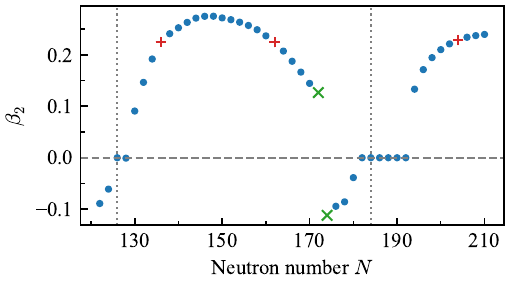}
    \caption{HFB ground-state quadrupole deformations of even-even plutonium 
    isotopes from $N=122$ to $N=210$. Isotopes denoted by red ``+'' markers at 
    $N=134$, $162$, and $204$ are shown in the right panel of 
    Fig.~\ref{fig:Pu_isotopes_photo}; isotopes marked by green crosses at 
    $N=172$ and $174$ are shown in Fig.~\ref{fig:Pu_266_268_photo}. Neutron 
    magic numbers $N=126$ and $184$ are marked by dotted vertical lines, 
    and $\beta_2=0$ is marked by a dashed horizontal line. }
    \label{fig:Pu_isotopes_deformations}
\end{figure}

To better analyze the responses along the isotopic chain, we first show in 
Fig.~\ref{fig:Pu_isotopes_deformations} the HFB ground-state quadrupole 
deformations of these Pu isotopes. We observe two transitions from an 
oblate to a prolate nuclear shape, one from neutron number $N=128$ to $130$, 
and the other from $N=180$ to $194$. There is also a transition from a prolate 
to an oblate shape at $N=172 \sim 174$. In the following we will discuss how 
these shape changes impact various multipole responses. 


\begin{figure*}
    \includegraphics[width=0.95\linewidth]{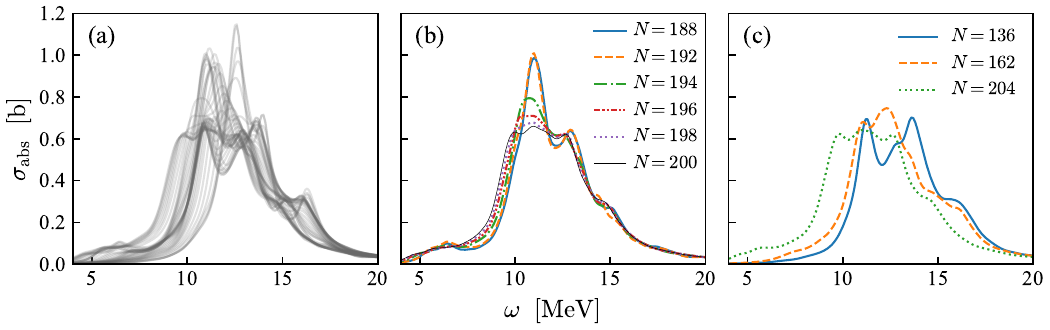}
    \caption{Panel (a): photoabsorption cross sections of even-even Pu isotopes 
    from the two-proton to the two-neutron dripline ($N=122$ to $210$), 
    calculated with the SLy4 parameterization and a width of $\Gamma=1$ MeV. Panel (b): 
    similar to panel (a) but for Pu isotopes around $N=194$, showing the effect resulting 
    from the onset of ground-state quadrupole deformation. Panel (c): similar to panel (a)  
    but for three different well-deformed, prolate Pu isotopes with 
    similar ground-state deformations $\beta_2 \approx 0.23$, denoted by red 
    plus markers in Fig.~\ref{fig:Pu_isotopes_deformations}. }
    \label{fig:Pu_isotopes_photo}
\end{figure*}

\subsubsection{$E1$ transitions}
\label{subsubsec:photo_Pu_isotopes}

We first present the $E1$ photoabsorption cross sections of all the even-even Pu 
isotopes in the left panel of Fig.~\ref{fig:Pu_isotopes_photo}, where we can 
clearly see how much the cross section varies as the neutron number $N$ changes. 
A detailed figure for the photoabsorption cross section of each plutonium 
isotope can be found in the Supplemental Material \cite{supp}. In the following we will 
closely examine the impact of the ground-state deformation on the $E1$ 
photoabsorption cross sections of Pu isotopes. 

The variations of the photoabsorption cross section around $N=194$, where a 
spherical-prolate shape transition occurs, are displayed in the middle panel of 
Fig.~\ref{fig:Pu_isotopes_photo}. As the neutron number increases, the cross 
section barely changes before the shape transition, but then the onset of 
quadrupole deformation significantly alters the cross section and makes it more 
fragmented, as the degeneracy on the angular-momentum projection of the 
external field is broken in a deformed nucleus. After the shape transition the 
cross section gradually becomes stabilized as the deformation $\beta_2$ also 
stabilizes. 

\begin{figure}[!htb]
    \includegraphics[width=0.95\linewidth]{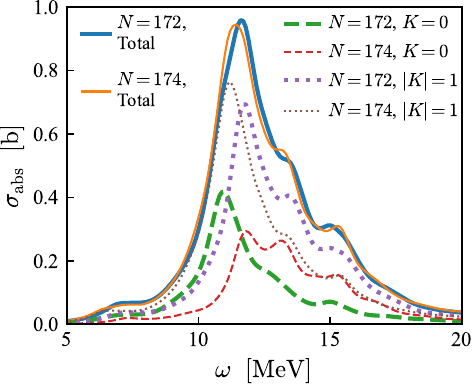}
    \caption{Photoabsorption cross sections of $^{266,268}$Pu, 
    as well as their $K=0$ and $|K|=1$ components, calculated with the SLy4 
    parameterization and a width of $\Gamma=1$ MeV. }
    \label{fig:Pu_266_268_photo}
\end{figure}

Besides the shape transitions from an oblate to a prolate shape, we also 
notice a prolate-oblate shape transition from $N=172$ to $174$ in 
Fig.~\ref{fig:Pu_isotopes_deformations}. As shown in 
Fig.~\ref{fig:Pu_266_268_photo}, this shape transition barely influences the 
total photoabsorption cross section, which is quite different from the 
spherical-prolate transition shown in Fig.~\ref{fig:Pu_isotopes_photo}. 
By decomposing the total cross section into $K=0$ 
and $|K|=1$ components, we can see in Fig.~\ref{fig:Pu_266_268_photo} that the 
two components change dramatically when the shape transition occurs, but their 
variations roughly compensate each other. When the prolate-oblate transition 
occurs, the nucleus shrinks in the $z$ direction but becomes more extended in 
$x$ and $y$ directions. Therefore, the effective $E1$ oscillators along the 
$z$ axis and in the perpendicular direction are basically swapped during the 
prolate-oblate transition, which leads to an almost unchanged total cross 
section. 

In order to better inspect the isotopic dependence, we show in the right panel 
of Fig.~\ref{fig:Pu_isotopes_photo} the photoabsorption cross sections of three 
Pu isotopes with similar deformations $\beta_2 \approx 0.23$ but different 
neutron numbers. We find that their cross sections have similar patterns, which 
can be attributed to their similar shapes. On the other hand, the cross-section 
curve in the right panel of Fig.~\ref{fig:Pu_isotopes_photo} shifts toward a 
lower energy as the neutron number increases, since the effective $E1$ 
oscillator length increases with the nuclear radius when the neutron number 
grows. 


\subsubsection{High-order multipole responses}
\label{subsubsec:high_order_Pu_isotopes}

\begin{figure*}
    \includegraphics[width=0.95\linewidth]{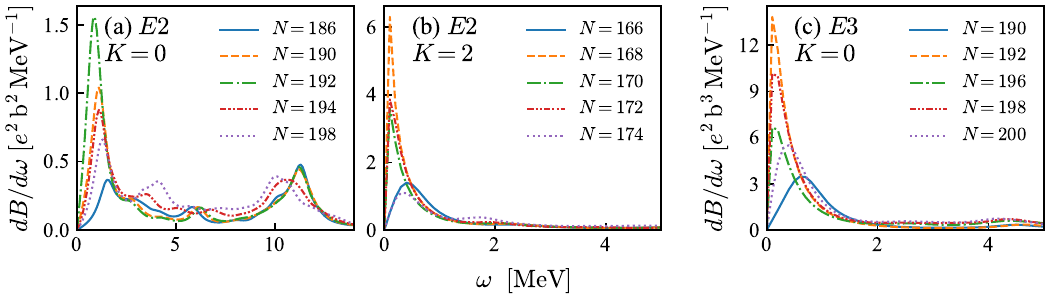}
    \caption{Panel (a): $K=0$ components of isoscalar $E2$ responses in Pu isotopes around $N=192$, 
    showing the onset of ground-state quadrupole deformation. 
    Panel (b): $K=2$ components of isoscalar $E2$ responses in Pu isotopes around $N=170$, 
    showing the onset and disappearance of ground-state $\gamma$ softness. 
    Panel (c): $K=0$ components of isoscalar $E3$ responses in Pu isotopes around $N=196$, 
    showing the onset and disappearance of ground-state octupole softness. 
    Responses in all the panels are calculated with the SLy4 parameterization 
    and a width of $\Gamma=1$ MeV. }
    \label{fig:Pu_isotopes_E2E3}
\end{figure*}

High-order responses in even-even plutonium isotopes have features similar to 
the $E1$ photoabsorption cross sections presented in 
Sec.~\ref{subsubsec:photo_Pu_isotopes}. Here we focus on the modes near 
$\omega=0$ for isoscalar $E2$ and $E3$ responses and discuss their 
relations with corresponding HFB ground-state multipole moments. 

Figure \ref{fig:Pu_isotopes_E2E3} shows the low-energy responses for the 
isoscalar $E2$ ($K=0$ and $2$) and isoscalar $E3$ ($K=0$) transitions 
in Pu isotopes that exhibit strong transition strengths near 
$\omega=0$. We do not show the $K=1$ component of the isoscalar $E2$ response 
because the rotational spurious mode in a deformed nucleus always gives a 
zero-energy peak in it. For the $K=0$ component of the $E2$ response presented in the
left panel of Fig.~\ref{fig:Pu_isotopes_E2E3}, the low-energy mode shifts 
leftward and becomes stronger as the neutron number increases to $N=192$ 
(before the spherical-prolate shape transition). This indicates that the 
quadrupole-deformed HFB minimum becomes lower in energy and closer to the 
energy of the spherical ground state. After the spherical-prolate shape 
transition, the low-energy mode becomes weaker and moves away from $\omega=0$. 

For the $K=2$ component of the isoscalar $E2$ response shown in the middle 
panel of Fig.~\ref{fig:Pu_isotopes_E2E3}, we observe similar strong low-energy 
peaks around $N=170$, just before the prolate-oblate shape transition. For 
these Pu isotopes, the energies of prolate and oblate HFB minima are 
getting closer to each other as the neutron number increases to $N=174$. We 
thus expect that these isotopes are $\gamma$-soft and tend to have 
triaxial-deformed ground states, which results in the low-energy modes seen in 
the middle panel of Fig.~\ref{fig:Pu_isotopes_E2E3}. 
For the isotopes of $N=168 \sim 172$, we also calculate their isoscalar $E2$ ($K=2$) 
responses along the imaginary axis in the complex $\omega$ plane and find poles
with imaginary QRPA energies, which indicates that their axially-deformed HFB 
states are not true HFB minima \cite{ring2004nuclear}. 

To verify this, we perform constrained $(Q_{20},Q_{22})$ HFB calculations 
with the triaxial code {\sc{hfodd}} \cite{schunck2017solution} for the isotopes of 
$N=166 \sim 174$. The resulting potential energy surfaces (PESs) are shown in the 
Supplemental Material \cite{supp}. We find that the PES for $N=166$ 
shows a well-pronounced axial minimum, which becomes weakly 
triaxial as the neutron number rises to $N=172$. For $N=174$, the PES
exhibits prolate-oblate shape coexistence, i.e., the energies of prolate and oblate 
minima are very close to one another.

A similar phenomenon can also be observed in the right panel of 
Fig.~\ref{fig:Pu_isotopes_E2E3} for the isoscalar $E3$ ($K=0$) responses around $N=196$. 
A strong low-energy $E3$ peak appears during the spherical-prolate shape transition, 
where Pu isotopes are known to be octupole-soft and can have pear-shaped ground states 
\cite{cao2020landscape}. For the isotopes of $N=192 \sim 198$, we do find  
the telltale imaginary poles in the $E3$ responses suggesting octupole softness. 
We recall that the HFB ground state on which the responses of 
Fig.~\ref{fig:Pu_isotopes_E2E3} are computed are reflection symmetric: when 
allowing the HFB solution to break the reflection symmetry, we find that the 
ground states of these isotopes become octupole-deformed and the resulting $E3$ 
responses no longer show the peaks seen in the panel (c) of Fig.~\ref{fig:Pu_isotopes_E2E3}. 

In summary, Fig.~\ref{fig:Pu_isotopes_E2E3} demonstrates that low-energy FAM responses  
are good diagnostic tools for ground-state multipole deformations and 
PES softness that may not be probed by the HFB solver; finding a pole on the imaginary 
$\omega$ axis can further confirm the softness. These indicators have been employed in 
Refs.~\cite{minhloc2023origin,leanh2024landscape} to explore the nuclear deformation 
softness with spherical QRPA calculations. Conversely, our results also show that  
accurate FAM descriptions of low-energy responses demand that we find the true HFB minimum. 


\section{Conclusions}
\label{sec:conclusion}

In this article, we developed an efficient FAM code to perform large-scale 
calculations for electromagnetic multipole responses in even-even, odd-$A$ and 
odd-odd heavy, deformed nuclei. We performed a careful analysis of the 
uncertainties of multipole responses induced by the parametrization of 
the Skyrme EDF, the calibration of the EDF (=statistical), and different 
configurations available for blocking in odd-mass and odd-odd nuclei. 

We find that the calculated photoabsorption cross sections of fissioning nuclei 
with standard Skyrme functionals are in good agreement with the reference 
database \cite{goriely2019reference}. In both spherical and deformed 
nuclei, the difference in the linear response between different 
parameterizations of the Skyrme EDF (systematic uncertainty) is much larger 
than the statistical uncertainty propagated from the Bayesian posterior 
distribution of Skyrme parameters (at least in the case of HFB1), 
which can be partly captured by the sum rule. Furthermore, 
in odd-mass and odd-odd nuclei the uncertainty of the response caused by 
different choices of blocking configurations is minor when compared with the 
systematic uncertainty. Importantly, even when the systematic uncertainty is relatively large, 
the overall structure of the multipole response is not impacted too much by the EDF parameterization. 
We confirm that information from electromagnetic responses could greatly help constrain isovector 
nuclear-matter properties in future Skyrme-EDF fits. 

As the first step toward global studies of electromagnetic transitions through 
the nuclear landscape, we calculated the multipole responses in even-even Pu 
isotopes from the proton to the neutron dripline. As the neutron number increases, 
the response varies dramatically when a transition from an oblate to a prolate 
ground-state shape happens, and it changes slowly after the transition as the 
nuclear shape stabilizes. An exception to this relation between the ground-state 
deformation and response occurs during the prolate-oblate shape transition, where the 
total photoabsorption cross section stays stable while the ground-state shape 
abruptly jumps from a prolate to an oblate minima. This can be understood by 
the swap of effective oscillator lengths along and perpendicular to the $z$ 
axis in the intrinsic frame, causing cancellation between the variations of 
different $K$ components. 

The infrastructure developed in this paper, together with the clarifications of 
the formalism and the analysis of uncertainties, is the first step toward 
large-scale calculations for nuclear responses over the entire chart of 
isotopes, which can provide crucial information for reaction and fission models as 
well as astrophysical studies.



\appendix
\section{Properties of FAM amplitudes}
\label{app:FAM_amplitudes}

We rewrite Eqs.~\eqref{eq:TD_qp_op} and \eqref{eq:XYPQ_def} as 
\begin{equation}\label{eq:transform_TDqp_staticqp}
    \begin{pmatrix}
        \beta(t) \\
        \beta^\dagger(t)
    \end{pmatrix} 
    =\mathbb{C}(t)\mathbb{W}(t)
    \begin{pmatrix}
        \beta \\ \beta^\dagger
    \end{pmatrix},
\end{equation}
where 
\begin{equation}
    \mathbb{C}_{\mu\nu}(t) = \begin{pmatrix}
        e^{iE_\mu t}\delta_{\mu\nu} & 0 \\
        0 & e^{-iE_\mu t}\delta_{\mu\nu}, 
    \end{pmatrix}
\end{equation}
and 
\begin{equation}
    \mathbb{W}(t) = \mathbbm{1} + \eta\delta\mathbb{W}^{(+)}(\omega) e^{-i \omega t} 
    + \eta\delta\mathbb{W}^{(-)}(\omega) e^{i \omega t}, 
\end{equation}
where 
\begin{subequations}
    \begin{align}
        \delta\mathbb{W}^{(+)}(\omega)&=
        \begin{pmatrix}
            P(\omega) & X^T(\omega) \\
            Y^T(\omega) & Q(\omega) 
        \end{pmatrix}, \\
        \delta\mathbb{W}^{(-)}(\omega)&=
        \begin{pmatrix}
            Q^*(\omega) & Y^\dagger(\omega) \\
            X^\dagger(\omega) & P^*(\omega) 
        \end{pmatrix}.
    \end{align}
\end{subequations}
The transformation matrix $\mathbb{C}(t)\mathbb{W}(t)$ in Eq.~\eqref{eq:transform_TDqp_staticqp} must be unitary, 
which means $\mathbb{W}(t)$ is unitary, i.e., 
$\mathbb{W}^\dagger(t) \mathbb{W}(t) = \mathbbm{1}$. 
We thus need
\begin{equation}
    \delta\mathbb{W}^{(+)\dagger}(\omega) + \delta\mathbb{W}^{(-)}(\omega) = 0
\end{equation}
to ensure the unitarity of $\mathbb{W}(t)$ up to the first order of $\eta$. 
This yields 
\begin{equation}\label{eq:XYPQ_properties}
    X =- X^T, \ 
    Y =- Y^T, \ 
    Q =- P^T. 
\end{equation}


\section{Induced FAM quantities and pairing cutoff}
\label{app:induced_quantities}

To calculate the induced densities and mean fields in the FAM,
we first write the time-dependent quasiparticle operator $\beta_\mu(t)$ 
in the s.p.\ basis: 
\begin{equation}
    \beta_\mu(t) = \sum_k \left[ U^*_{k\mu}(t) c_k + V^*_{k\mu}(t) c^\dagger_k \right] e^{iE_\mu t}.
\end{equation}
with
\begin{subequations}\label{eq:TD_U_V}
    \begin{align}
        U(t) &= U + \eta \delta U^{(+)}(\omega) e^{-i\omega t} + \eta \delta U^{(-)}(\omega) e^{i\omega t}, \\
        V(t) &= V + \eta \delta V^{(+)}(\omega) e^{-i\omega t} + \eta \delta V^{(-)}(\omega) e^{i\omega t},
    \end{align}
\end{subequations}
where $U$ and $V$ define the static Bogoliubov transformation  
\begin{equation}
    \beta_\mu = \sum_k \left[ U^*_{k\mu} c_k + V^*_{k\mu} c^\dagger_k \right].
\end{equation}
Based on Eqs.~\eqref{eq:TD_qp_op} and \eqref{eq:XYPQ_def}, we can express $\delta U^{(\pm)}(t)$ and $\delta V^{(\pm)}(t)$ in terms of FAM amplitudes: 
\begin{subequations}\label{eq:induced_U_V}
    \begin{align}
        \delta U_{k\mu}^{(+)}(\omega) &= \Big[ U Q^T(\omega) +       V^* Y(\omega)         \Big]_{k\mu}, \\
        \delta U_{k\mu}^{(-)}(\omega) &= \Big[ U P^\dagger(\omega) + V^* X^*(\omega) \Big]_{k\mu}, \\
        \delta V_{k\mu}^{(+)}(\omega) &= \Big[ V Q^T(\omega) +       U^* Y(\omega)         \Big]_{k\mu}, \\
        \delta V_{k\mu}^{(-)}(\omega) &= \Big[ V P^\dagger(\omega) + U^* X^*(\omega) \Big]_{k\mu}.
    \end{align}
\end{subequations}

With the help of time-dependent quasiparticle spinors $U(t)$ and $V(t)$,
we can obtain the time-dependent density matrix and pairing tensor as \cite{goodman1981finitetemperature}
\begin{subequations}\label{eq:TD_rho_kappa}
    \begin{align}
        \rho_{kl}(t)   &= \sum_{\mu} \Big[ f_\mu U_{k\mu}(t) U^*_{l\mu}(t) + g_\mu V^*_{k\mu}(t) V_{l\mu}(t) \Big], \\
        \kappa_{kl}(t) &= \sum_{\mu} \Big[ f_\mu U_{k\mu}(t) V^*_{l\mu}(t) + g_\mu V^*_{k\mu}(t) U_{l\mu}(t) \Big],
    \end{align}
\end{subequations}
where 
\begin{equation}\label{eq:define_g}
    g_\mu = 1-f_\mu. 
\end{equation} 
Besides, $\rho(t)$ and $\kappa(t)$ oscillate in the same manner as the external field $F(t)$ 
in Eq.~\eqref{eq:F_oscillate}:
\begin{subequations}\label{eq:rho_kappa_oscillate}
    \begin{align}
        \rho(t)   & = \rho_0  + \eta \Big[ \delta \rho(\omega) e^{-i\omega t} + \delta \rho^\dagger(\omega) e^{i\omega t}        \Big], \\
        \kappa(t) &= \kappa_0 + \eta \Big[ \delta \kappa^{(+)}(\omega) e^{-i\omega t} + \delta \kappa^{(-)}(\omega) e^{i\omega t}\Big].
    \end{align}
\end{subequations}

Substituting Eqs.~\eqref{eq:TD_U_V}, \eqref{eq:induced_U_V}, \eqref{eq:TD_rho_kappa} into \eqref{eq:rho_kappa_oscillate}
and keeping terms linear in $\eta$, we can obtain the induced density matrix and pairing tensor: 
\begin{widetext}
\begin{subequations}\label{eq:rho_kappa_from_PQXY_explicit}
    \begin{align}
        \delta\rho_{kl}(\omega) &= \sum_{\mu\nu} \Big[
             \left(f_\mu-f_\nu\right) U_{k\mu}   P_{\mu\nu} U^*_{l\nu}
            +\left(g_\mu-g_\nu\right) V^*_{k\mu} Q_{\mu\nu} V_{l\nu}
            +\left(g_\nu-f_\mu\right) U_{k\mu}   X_{\mu\nu} V_{l\nu}
            -\left(g_\mu-f_\nu\right) V^*_{k\mu} Y_{\mu\nu} U^*_{l\nu} \Big], \\
        \delta\kappa^{(+)}_{kl}(\omega) &= \sum_{\mu\nu} \Big[
             \left(f_\mu-f_\nu\right) U_{k\mu}   P_{\mu\nu} V^*_{l\nu}
            +\left(g_\mu-g_\nu\right) V^*_{k\mu} Q_{\mu\nu} U_{l\nu}
            +\left(g_\nu-f_\mu\right) U_{k\mu}   X_{\mu\nu} U_{l\nu}
            -\left(g_\mu-f_\nu\right) V^*_{k\mu} Y_{\mu\nu} V^*_{l\nu} \Big], \\
        \delta\kappa^{(-)}_{kl}(\omega) &= \sum_{\mu\nu} \Big[
             \left(f_\mu-f_\nu\right) V^*_{l\mu} P^*_{\mu\nu} U_{k\nu}
            +\left(g_\mu-g_\nu\right) U_{l\mu}   Q^*_{\mu\nu} V^*_{k\nu}
            +\left(g_\nu-f_\mu\right) V^*_{l\mu} X^*_{\mu\nu} V^*_{k\nu}
            -\left(g_\mu-f_\nu\right) U_{l\mu}   Y^*_{\mu\nu} U_{k\nu} \Big].
    \end{align}
\end{subequations}
\end{widetext}
With the help of Eqs.~\eqref{eq:XYPQ_properties} and \eqref{eq:define_g}, one 
can show that $\delta\kappa^{(\pm)}(\omega)$ is antisymmetric while 
$\delta\rho(\omega)$ is not necessarily Hermitian. 

In an even-even nucleus at zero temperature, we have $f_\mu = 0$ and $g_\mu = 1$, 
so the amplitudes $P$ and $Q$ do not contribute in 
Eq.~\eqref{eq:rho_kappa_from_PQXY_explicit}. In this case, we can thus exclude 
$P$ in the FAM equations \eqref{eq:FAM_equation_P}, which leads to the formalism 
given in Ref.~\cite{avogadro2011finite}. With the expressions of $\delta\rho$ 
and $\delta\kappa^{(\pm)}$, one can then compute the induced mean field 
$\delta H$ using equations presented in \cite{avogadro2011finite}. All the 
matrices involved in the FAM calculation have a dimension of $N_{2\mathrm{qp}}$; 
they are much smaller than the QRPA matrix and can thus be efficiently 
constructed. 

In this work we adopt a zero-range pairing force, which is known to cause
divergence as the model space increases \cite{bulgac2002renormalization,
borycki2006pairing,schunck2019energy}. In static HFB calculations, the
divergence can be mitigated by introducing a quasiparticle cutoff, i.e., all the
quasiparticles with energies above a given threshold $E_{\rm cut}$ are
discarded in the calculations of densities. In the FAM we utilize a similar
cutoff recipe \cite{li2024numerical}: the summations in
Eq.~\eqref{eq:TD_rho_kappa} only run over quasiparticle indices $\mu$ within
the pairing window of the \textit{static} HFB solution, which is equivalent to
replacing $f_\mu$ and $g_\mu$ in Eq.~\eqref{eq:rho_kappa_from_PQXY_explicit}
with
\begin{equation}\label{eq:f_g_cutoff}
    \xi_\mu^{\rm cut} = \begin{cases}
        \xi_\mu & E_\mu < E_{\rm cut} \\
        0 & \text{otherwise}
    \end{cases}, 
    \text{ for } \xi = f \text{ or } g. 
\end{equation}

One negative consequence of this cutoff method is that the amplitude $Q$ now 
contributes to the induced densities \eqref{eq:rho_kappa_from_PQXY_explicit} in
an even-even nucleus, because $g_\mu-g_\nu \neq 0$ when $\mu$ is inside the
pairing window while $\nu$ is outside. To avoid this issue we force
$g_\mu - g_\nu = f_\nu - f_\mu$ in our calculations. Furthermore, the pairing
cutoff breaks the antisymmetry of the induced pairing tensor
$\delta\kappa^{(\pm)}$ since Eq.~\eqref{eq:define_g}, which is employed to
prove the antisymmetry, is not consistent with the cutoff \eqref{eq:f_g_cutoff}.
These inconsistencies can be removed by implementing the renormalization of the
pairing force \cite{bulgac2002renormalization} or by using finite-range
pairing forces.


\section{Energy-weighted sum rules for electric multipole transitions}
\label{app:sum_rule}

In this appendix we follow the procedure presented in 
Ref.~\cite{lipparini1989sum} to evaluate Eq.~\eqref{eq:m1_def} for the Skyrme 
EDF. For the electric multipole operator $Q_{LK}$ 
\eqref{eq:electric_multipole_op}, we have 
\begin{equation}\label{eq:m1_T_electric_op}
    m_1^T(Q) = \frac{e^2 \hbar^2}{2m} \int |\vec{\nabla} f(\vec{r})|^2 \!
    \left[ q_n^2 \rho_n(\vec{r}) \!+\! q_p^2 \rho_p(\vec{r}) \right] d\vec{r}, 
\end{equation}
where $f(\vec{r}) = r^L Y_{LK}(\Omega)$. Assuming a gauge-invariant Skyrme EDF, 
we can write the enhancement factor $\kappa(Q)$ as 
\begin{equation}\label{eq:kappa_electric_op}
    \kappa(Q) = \frac{e^2\! \left( q_n - q_p \right)^2\! \left( C^\tau_0 - C^\tau_1 \right)}{m_1^T(Q)}
    \int |\vec{\nabla} f|^2 \rho_n(\vec{r}) \rho_p(\vec{r}) d\vec{r},  
\end{equation}
where $C^\tau_t$ is the coupling constant of the term $\rho_t \tau_t$ in the 
Skyrme EDF. 

Using standard formulas for the gradient of spherical harmonics in a spherical 
tensor basis \cite{varshalovich1988quantum}, we find
\begin{equation}
    \begin{aligned}
        \left(\vec{\nabla} f \right) \cdot \vec{e}_\mu =& 
        (-1)^{L+K} \sqrt{L}(2L+1) \begin{pmatrix}
            L\!-\!1    & 1    & L \\
            K\!-\!\mu  & \mu  & -\!K
        \end{pmatrix} \\
        & \times r^{L-1} Y_{L-1, K-\mu}(\Omega),\quad \mu=0,\pm 1, 
    \end{aligned}
\end{equation}
which involves the 3$j$ symbol $\begin{pmatrix}
    j_1 & j_2 & j_3 \\
    m_1 & m_2 & m_3
\end{pmatrix}$. 
Then we can write $\left| \vec{\nabla} f \right|^2$ as 
\begin{equation}\label{eq:nabla_f_sqr_in_Y}
    \left| \vec{\nabla} f(\vec{r}) \right|^2 = r^{2L-2}
    \sum_{J=0,2,\cdots}^{2L-2} \mathcal{C}_{LK}^{(J)} Y_{J 0}(\Omega), 
\end{equation}
where 
\begin{equation}\label{eq:C_LK^J}
    \begin{aligned}
        \mathcal{C}_{LK}^{(J)} &=  L(2L+1)^2 (2L-1) \sqrt{\frac{2J+1}{4\pi}} 
        \begin{pmatrix}
            L\!-\!1    & L\!-\!1    & J \\
            0  & 0  & 0
        \end{pmatrix} \\
        & \times \sum_{\mu=-1}^{1} (-1)^M
        \begin{pmatrix}
            L\!-\!1    & 1    & L \\
            M  & \mu  & -\!K
        \end{pmatrix}^2 
        \begin{pmatrix}
            L\!-\!1    & L\!-\!1    & J \\
            -M  & M  & 0
        \end{pmatrix}, 
    \end{aligned}
\end{equation}
where $M=K-\mu$, and 
the first and third 3$j$ coefficients come from the product of $Y_{L-1, K-\mu}^*$ and $Y_{L-1, K-\mu}$
\cite{varshalovich1988quantum}. One can 
see that $\left| \vec{\nabla} f \right|^2$ always has even parity, so the sum 
over $J$ in \eqref{eq:nabla_f_sqr_in_Y} involves only even numbers. For example, 
we give below the explicit expressions of 
$\left| \vec{\nabla} f(\vec{r}) \right|^2$ for the $E2$ transition: 
\begin{subequations}\label{eq:nabla_f_2_sqr}
    \begin{equation}\label{eq:nabla_f_20_sqr}
        \begin{aligned}
            \Big| \vec{\nabla} \left[ r^2 Y_{20}(\Omega) \right] \Big|^2 &= 
            \sqrt{\frac{5}{\pi}} r^2 \left[ \sqrt{5}Y_{00}(\Omega) + Y_{20}(\Omega) \right] \\
            &= \frac{5}{4\pi} \left( 4z^2 + x^2 + y^2 \right), 
        \end{aligned}
    \end{equation}
    \begin{equation}
        \begin{aligned}
            \Big| \vec{\nabla} \left[ r^2 Y_{21}(\Omega) \right] \Big|^2 &= 
            \frac{1}{2}\sqrt{\frac{5}{\pi}} r^2 \left[ 2\sqrt{5}Y_{00}(\Omega) + Y_{20}(\Omega) \right] \\
            &= \frac{15}{8\pi} \left( 2z^2 + x^2 + y^2 \right), 
        \end{aligned}
    \end{equation}
    \begin{equation}\label{eq:nabla_f_22_sqr}
        \begin{aligned}
            \Big| \vec{\nabla} \left[ r^2 Y_{22}(\Omega) \right] \Big|^2 &= 
            \sqrt{\frac{5}{\pi}} r^2 \left[ \sqrt{5}Y_{00}(\Omega) - Y_{20}(\Omega) \right]  \\
            &= \frac{15}{4\pi} \left( x^2 + y^2 \right), 
        \end{aligned}
    \end{equation}
\end{subequations}
where Eq.~\eqref{eq:nabla_f_20_sqr} agrees with the expression derived in 
Ref.~\cite{hinohara2015complexenergy}. With Eq.~\eqref{eq:nabla_f_sqr_in_Y}, the 
energy-weighted sum rule \eqref{eq:m1_T_electric_op} for the electric multipole 
operator $Q_{LK}$ can be expressed as a function of the radial moment 
$\langle r^{2L-2} \rangle$ and multipole deformations $\beta_2$, $\beta_4$, ..., 
$\beta_{2L-2}$ of the HFB ground state \cite{hinohara2015complexenergy}. 
Therefore, constraining the EDF parameters with measurements on high-order 
radial moments \cite{reinhard2020charge} is expected to improve the calculations of electric multipole 
responses. 


\begin{acknowledgments}
Discussions with Antonio Bjelcic, Jonathan Engel, Kyle Godbey, Nobuo Hinohara, Eunjin In, 
Bui Minh Loc, Witold Nazarewicz, Gregory Potel, and Marc Verriere are gratefully acknowledged.
This work was partly performed under the auspices of the U.S.\ Department of 
Energy by the Lawrence Livermore National Laboratory under Contract 
DE-AC52-07NA27344. 
Work at Los Alamos National Laboratory was carried out under the auspices of 
the National Nuclear Security Administration of the U.S.\ Department of Energy 
under Contract No.\ 89233218CNA000001.
This material is based upon work supported by the U.S.\ Department of Energy, 
Office of Science, Office of Advanced Scientific Computing Research and Office of 
Nuclear Physics, Scientific Discovery through Advanced Computing (SciDAC) program.
Computing support for this work came from the Lawrence 
Livermore National Laboratory Institutional Computing Grand Challenge 
program.
\end{acknowledgments} 

\input{main.bbl}
\end{document}


\preprint{LLNL-JRNL-866328}
\preprint{LA-UR-24-27051}

\begin{CJK*}{UTF8}{gbsn}

\title{Supplemental material: Multipole response in fissioning nuclei and their uncertainties}

\author{Tong \surname{Li} (李通)}
\affiliation{Nuclear and Chemical Sciences Division, Lawrence Livermore National Laboratory, 
Livermore, CA 94551, USA}

\author{Nicolas \surname{Schunck}}
\affiliation{Nuclear and Chemical Sciences Division, Lawrence Livermore National Laboratory, 
Livermore, CA 94551, USA}

\author{Mike \surname{Grosskopf}}
\affiliation{Computer, Computational, and Statistical Sciences Division, Los Alamos National Laboratory, 
Los Alamos, NM 87545, USA}

\maketitle

\end{CJK*}


\section{Bayesian Posterior Distribution}

To perform Bayesian calibration for Skyrme parameters, we use the same 
approach as previous Bayesian calibration works \cite{mcdonnell2015uncertainty, 
higdon2015bayesian,schunck2020calibration}. 
First, 500 evaluations of the HFB1 model were performed for Skyrme 
parameters generated on a Latin hypercube within the bounds set in 
\cite{schunck2020calibration}. These evaluations were used to fit a 
multivariate Gaussian process (GP) surrogate model by maximizing the marginal 
likelihood of the GP, as described in \cite{schunck2020calibration}. 

\begin{figure}
    \includegraphics[width=\linewidth]{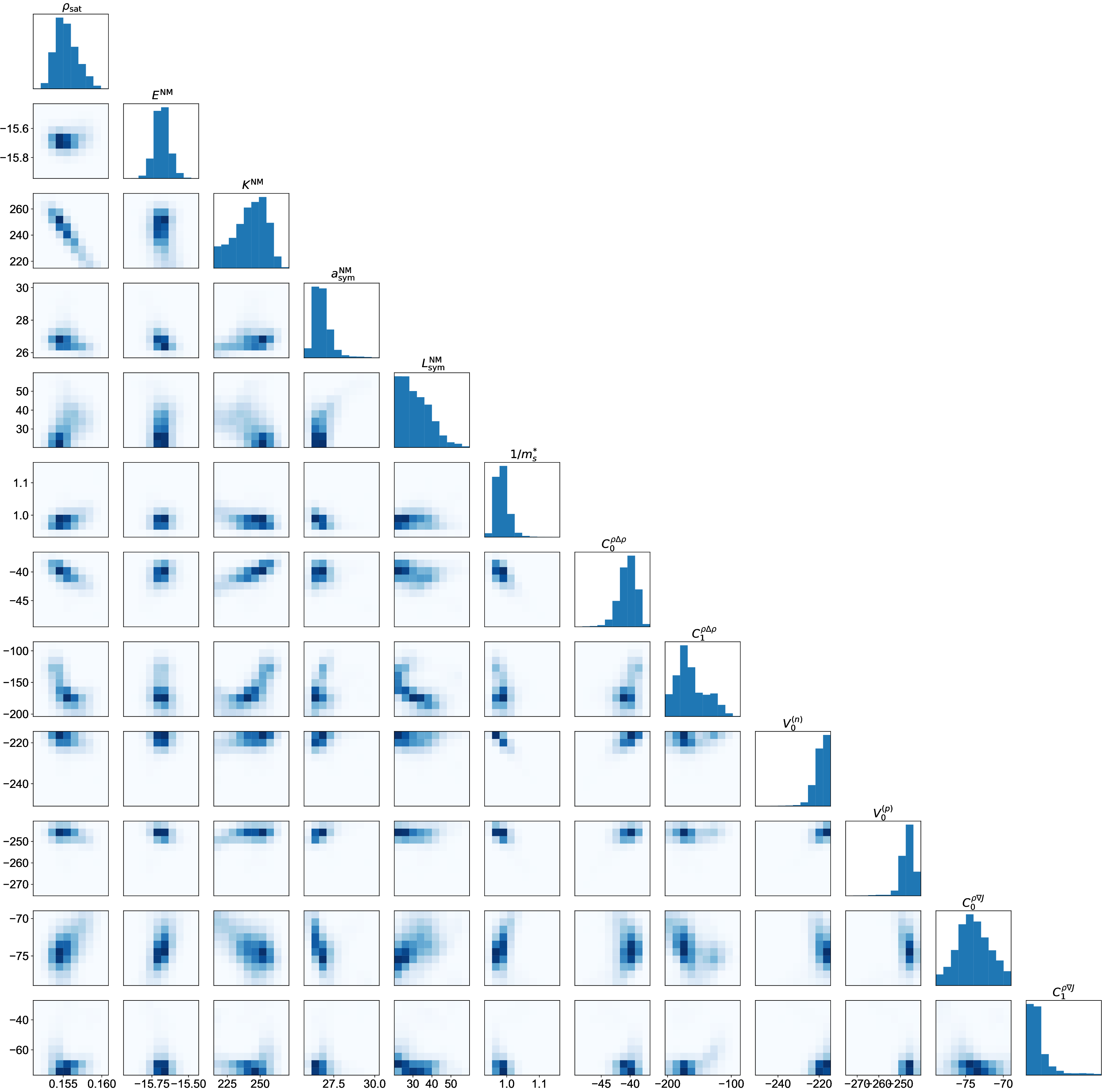}
    \caption{Univariate and bivariate histograms of posterior samples from the 
             Bayesian posterior distribution for Skyrme parameters.}
    \label{fig:posterior_histograms}
\end{figure}

\begin{figure}
    \includegraphics[width=\linewidth]{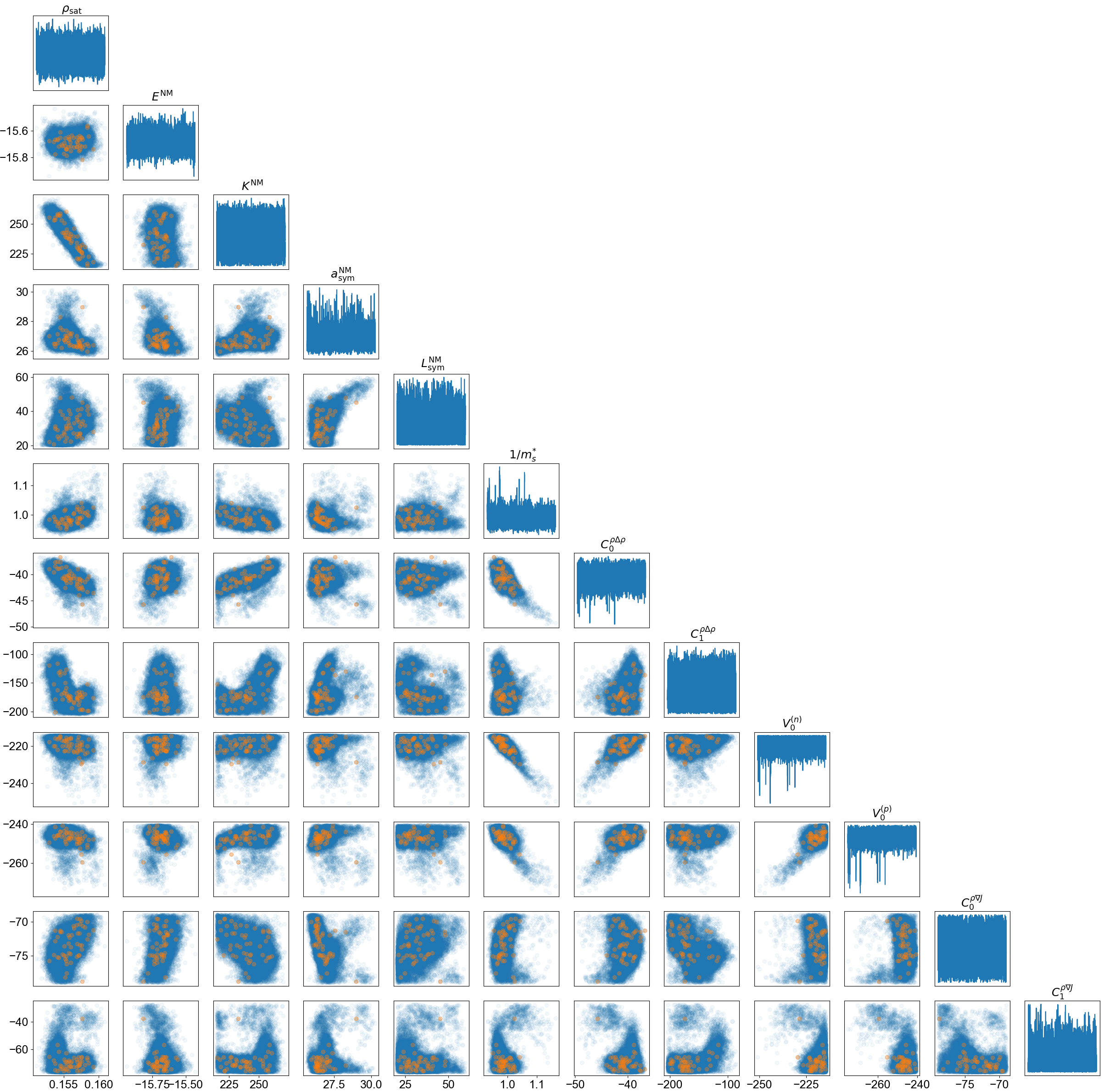}
    \caption{The off-diagonal panels show bivariate scatter plots of posterior 
             samples for Skyrme parameters, with the selected 50 samples inside the 
             95\% credible region highlighted in orange. The on-diagonal panels 
             show \emph{trace plots}, the sampled parameter values plotted 
             against the MCMC iteration, which are common MCMC diagnostics.}
    \label{fig:samples_overlaid}
\end{figure}

The fitted surrogate was implemented in the probabilistic programming language 
Stan, to fit the Bayesian calibration model with the No U-Turn Sampler (NUTS) 
\cite{carpenter2017stan}. Univariate and bivariate histograms of the posterior 
distribution are shown in Fig.~\ref{fig:posterior_histograms}. Parameters 
that are consistent with experimental observations are tightly constrained, with some 
cross-parameter correlation and complex structure. The off-diagonal panels in 
Fig.~\ref{fig:samples_overlaid} show the same posterior samples as a 
scatter plot; we randomly select 50 samples inside the 95\% credible region and 
use them to propagate the statistical uncertainty to FAM responses. The on-diagonal 
panels show common Markov chain Monte Carlo (MCMC) diagostic plots - \emph{trace plots}. Each plot 
concatenates four independent Markov chains and does not appear to have 
systematic structure or variations from chain to chain, consistent with 
well-mixed Markov chains that have converged to their stationary distribution.

\section{Pairing Strengths}

We use the following pairing interaction
\begin{equation}
V(\boldsymbol{r},\boldsymbol{r}') = 
V_{0}^q \left[ 1 - \frac{1}{2} \frac{\rho(\boldsymbol{r})}{\rho_c} \right] \delta(\boldsymbol{r}-\boldsymbol{r}').
\end{equation}
where $\rho(\boldsymbol{r})$ is the isoscalar density and $\rho_c = 0.16$ 
fm$^{-3}$. Table \ref{tab:pairing_strengths} lists the neutron ($q\equiv n$) 
and proton ($q\equiv p$) pairing strengths $V_{0}^q$ for each of the 
functionals considered in this work. The strengths were adjusted to fit the 
3-point odd-even mass staggering in $^{232}$Th.

\begin{table}[!htb]
    \caption{Pairing strengths $V_0^q$ in the unit of MeV $\cdot$ fm$^3$ for 
             various Skyrme EDFs. They are fitted to reproduce the three-point 
             odd-even mass staggering in $^{232}$Th. }
    \label{tab:pairing_strengths}
    \begin{ruledtabular}
    \begin{tabular}{ccccc}
            & SLy4 & SLy5 & SkM* & SkI3 \\
    \hline
    $V_0^n$ & -300.213 & -297.149 & -260.946 & -351.828 \\
    $V_0^p$ & -336.292 & -334.353 & -321.045 & -376.039 \\
    \end{tabular}
    \end{ruledtabular}
\end{table}


\section{Average Peak Energy}

\begin{figure}
    \includegraphics[width=0.95\linewidth]{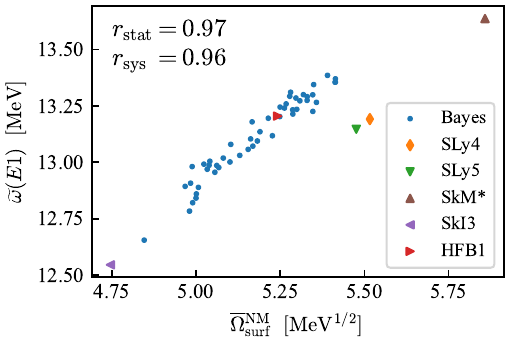}
    \caption{Similar to Fig.~3 in the main text, 
             but for $\widetilde{\omega}(E1)$ instead of $\overline{\omega}(E1)$}. 
    \label{fig:Pb208_avgtilde_omega}
\end{figure}

According to Sec.~IV B in the main text, 
we can use the sum rules $m_k(F)$ to define the average peak energy in two different ways 
\cite{bohigas1979sum}: 
\begin{equation}\label{eq:avg_E1_peak_pos}
    \overline{\omega}(F) = \frac{m_1(F)}{m_0(F)} \geq 
    \widetilde{\omega}(F) = \sqrt{\frac{m_1(F)}{m_{-1}(F)}}. 
\end{equation}
In the main text we have discussed the correlation between $\overline{\omega}(E1)$ 
and the nuclear-surface isovector oscillating frequency 
\begin{equation}\label{eq:frequency_NM_surf_isospin}
    \overline{\Omega}^{\rm NM}_{\rm surf} 
    = \sqrt{\left( a_{\rm sym}^{\rm NM} - \frac{L_{\rm sym}^{\rm NM}}{6} \right)
    M_v^{*-1}}. 
\end{equation}
Figure \ref{fig:Pb208_avgtilde_omega} shows the correlation between another average energy
$\widetilde{\omega}(E1)$ and $\overline{\Omega}^{\rm NM}_{\rm surf}$. 
The correlation shown in Fig.~\ref{fig:Pb208_avgtilde_omega} is slightly stronger 
than that displayed in Fig.~3 in the main text. 

\section{Photoabsorption Cross Sections and Multipole Responses}

\begin{figure}
    \includegraphics[width=\linewidth]{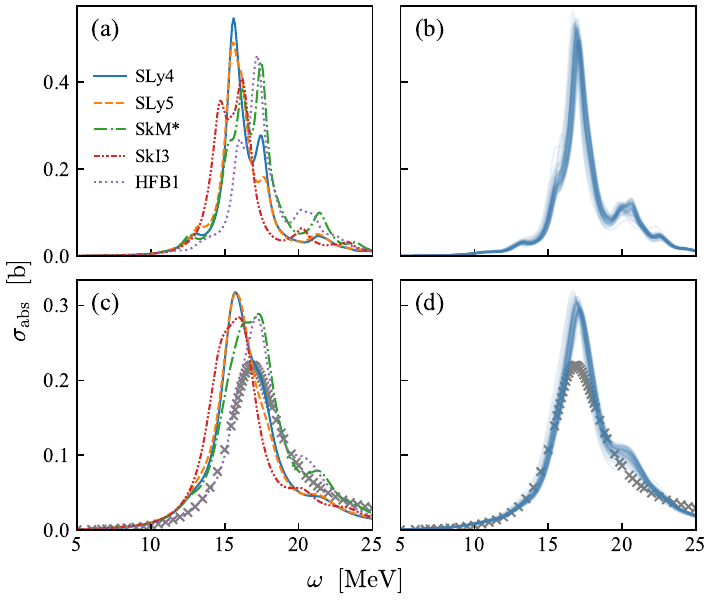}
    \caption{Panel (a): photoabsorption cross sections of $^{90}$Zr calculated with  
             different Skyrme parameterizations and a width of 
             $\Gamma = 1$ MeV. Panel (b): similar to panel (a) but calculated with 
             the 50 samples from the HFB1 posterior distribution. 
             Panels (c) and (d): Similar to panels 
             (a) and (b) but calculated with a width of $\Gamma = 2$ MeV. The 
             evaluated nuclear data \cite{goriely2019reference} are given by cross 
             markers. }
    \label{fig:Zr90_photo}
\end{figure}

Figure \ref{fig:Zr90_photo} shows the $E1$ photoabsorption cross sections of 
$^{90}$Zr, We see in Fig.~\ref{fig:Zr90_photo} that the systematic uncertainty 
is larger than the statistical one, which is consistent with the case of 
$^{208}$Pb discussed in the main text. 

\begin{figure}
    \includegraphics[width=0.95\linewidth]{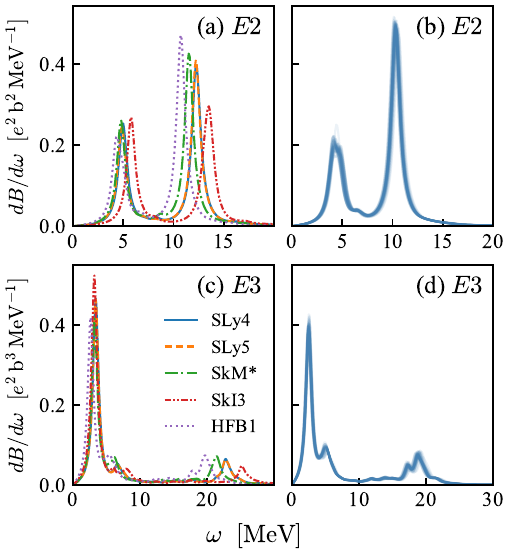}
    \caption{Panel (a): isoscalar $E2$ responses in $^{208}$Pb calculated with different 
             Skyrme EDFs and a width of $\Gamma=1$ MeV. 
             Panel (b): similar to panel (a) but calculated with the 50 samples 
             from the HFB1 posterior distribution. Panels (c) and (d): similar to panels 
             (a) and (b) but for isoscalar $E3$ responses. }
    \label{fig:Pb208_E2E3_isoscalar}
\end{figure}

\begin{figure}
    \includegraphics[width=0.95\linewidth]{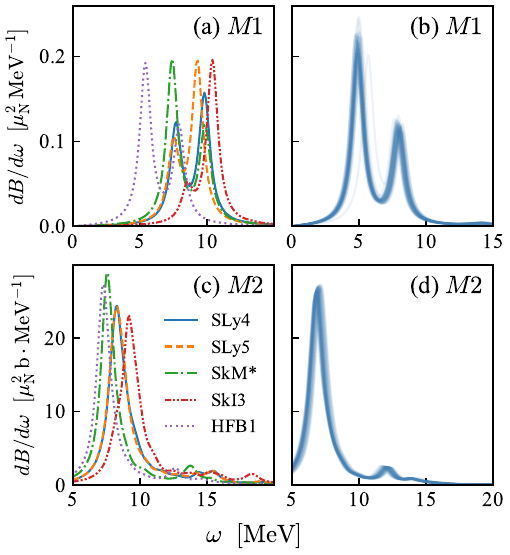}
    \caption{Panel (a): isoscalar $M1$ responses in $^{208}$Pb calculated with different 
             Skyrme EDFs and a width of $\Gamma=1$ MeV. 
             Panel (b): similar to panel (a) but calculated with the 50 samples 
             from the HFB1 posterior distribution. Panels (c) and (d): similar to panels  
             (a) and (b) but for isoscalar $M2$ responses. }
    \label{fig:Pb208_M1M2_isoscalar}
\end{figure}

Figure \ref{fig:Pb208_E2E3_isoscalar} presents the isoscalar $E2$ and $E3$ 
responses in $^{208}$Pb, calculated with different Skyrme parameterizations, 
while Fig.~\ref{fig:Pb208_M1M2_isoscalar} shows the isoscalar $M1$ and $M2$ 
responses. Both figures confirm the conclusions given in the main text. 

\begin{figure*}
    \includegraphics[width=0.95\linewidth]{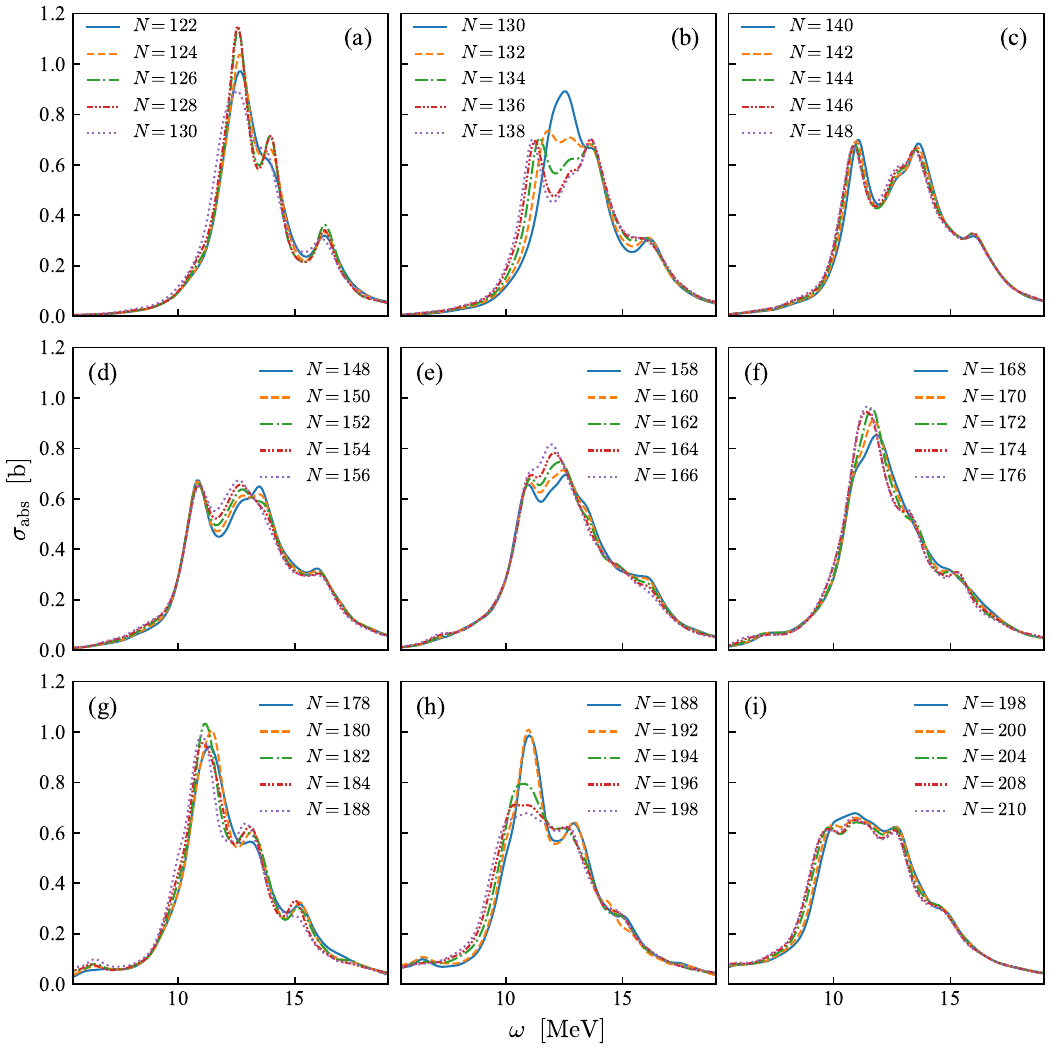}
    \caption{$E1$ photoabsorption cross sections of even-even Pu isotopes from 
             the two-proton to the two-neutron dripline ($N=122$ to $210$), 
             calculated with the SLy4 parameterization and a width of $\Gamma=1$ MeV. 
             Some isotopes are omitted as their cross sections are close to those 
             of nearby isotopes. }
    \label{fig:Pu_isotopes_all_photo}
\end{figure*}

Figure \ref{fig:Pu_isotopes_all_photo} gives the $E1$ photoabsorption cross 
sections of even-even Pu isotopes from the two-proton to the two-neutron dripline, 
calculated with the SLy4 parameterization. We find that the cross section 
varies significantly at shape transitions from oblate to spherical or spherical 
to prolate deformation ($N= 124 \sim 130$, $N= 180 \sim 184$, and 
$N= 190 \sim 194$), but it changes slowly in between. The prolate-oblate shape 
transition ($N=172 \sim 174$) barely impacts the photoabsoprtion cross section; 
see the main text for discussions. 

\section{Blocking Configurations}

Tables \ref{tab:Np238_blocking_n} and \ref{tab:Np238_blocking_n} list the
candidates for neutron and proton blocked states in $^{238}$Np, respectively, 
sorted by their quasiparticle energies in ascending order. 

\begin{table}[H]
    \caption{Candidates for neutron blocked orbits in $^{238}$Np within a 2 MeV 
             quasiparticle-energy window, labeled by asymptotic quantum numbers 
             $[N n_z \Lambda] \Omega^\pi$. They are given by the HFB 
             calculation of $^{238}$U with the SLy4 parameterization and sorted 
             by their quasiparticle energies in ascending order. 
             The blocking configuration for the ground state is highlighted in bold. }
    \label{tab:Np238_blocking_n}
    \begin{ruledtabular}
    \begin{tabular}{ccccc}
        $[ 6 2 2]5/2^+$ & $\boldsymbol{[7 4 3]7/2^-}$ & $[ 6 3 1]1/2^+$ & $[ 6 2 4]7/2^+$ & $[ 7 5 2]5/2^-$ \\
        $[ 6 3 3]5/2^+$ & $[ 7 3 4]9/2^-$ & $[ 6 1 3]7/2^+$ & $[ 7 7 0]1/2^-$ & $[ 6 2 0]1/2^+$ \\
        $[ 6 2 2]3/2^+$ & $[ 6 3 1]3/2^+$ & & & \\
    \end{tabular}
    \end{ruledtabular}
\end{table}

\begin{table}[H]
    \caption{Similar to Table \ref{tab:Np238_blocking_n} but for proton blocked orbits. }
    \label{tab:Np238_blocking_p}
    \begin{ruledtabular}
    \begin{tabular}{ccccc}
        $\boldsymbol{[5 2 3]5/2^-}$ & $[ 6 5 1]3/2^+$  & $[ 6 4 2]5/2^+$  & $[ 4 0 0]1/2^+$  & $[ 5 2 1]3/2^-$ \\
        $[ 5 3 0]1/2^-$ & $[ 4 0 0]1/2^+$  & $[ 5 0 5]11/2^-$ & $[ 4 0 2]3/2^+$  & $[ 5 3 2]3/2^-$ \\
        $[ 5 2 1]1/2^-$ & $[ 5 1 4]9/2^-$  & $[ 6 3 3]7/2^+$  & & \\
    \end{tabular}
    \end{ruledtabular}
\end{table}


\section{Potential Energy Surfaces in Neutron-rich Plutonium Isotopes}

\begin{figure*}
    \includegraphics[width=0.48\linewidth]{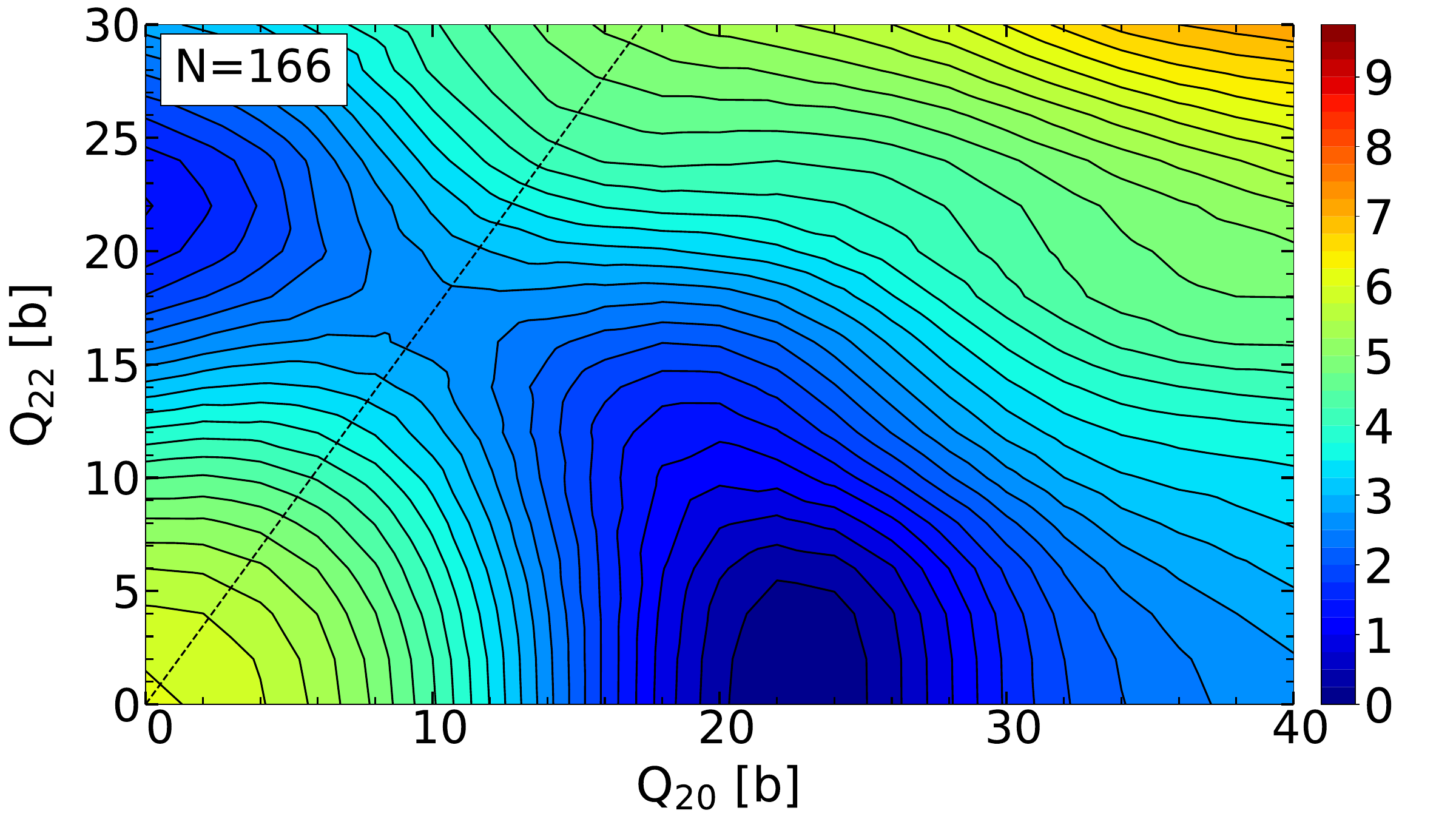}
    \includegraphics[width=0.48\linewidth]{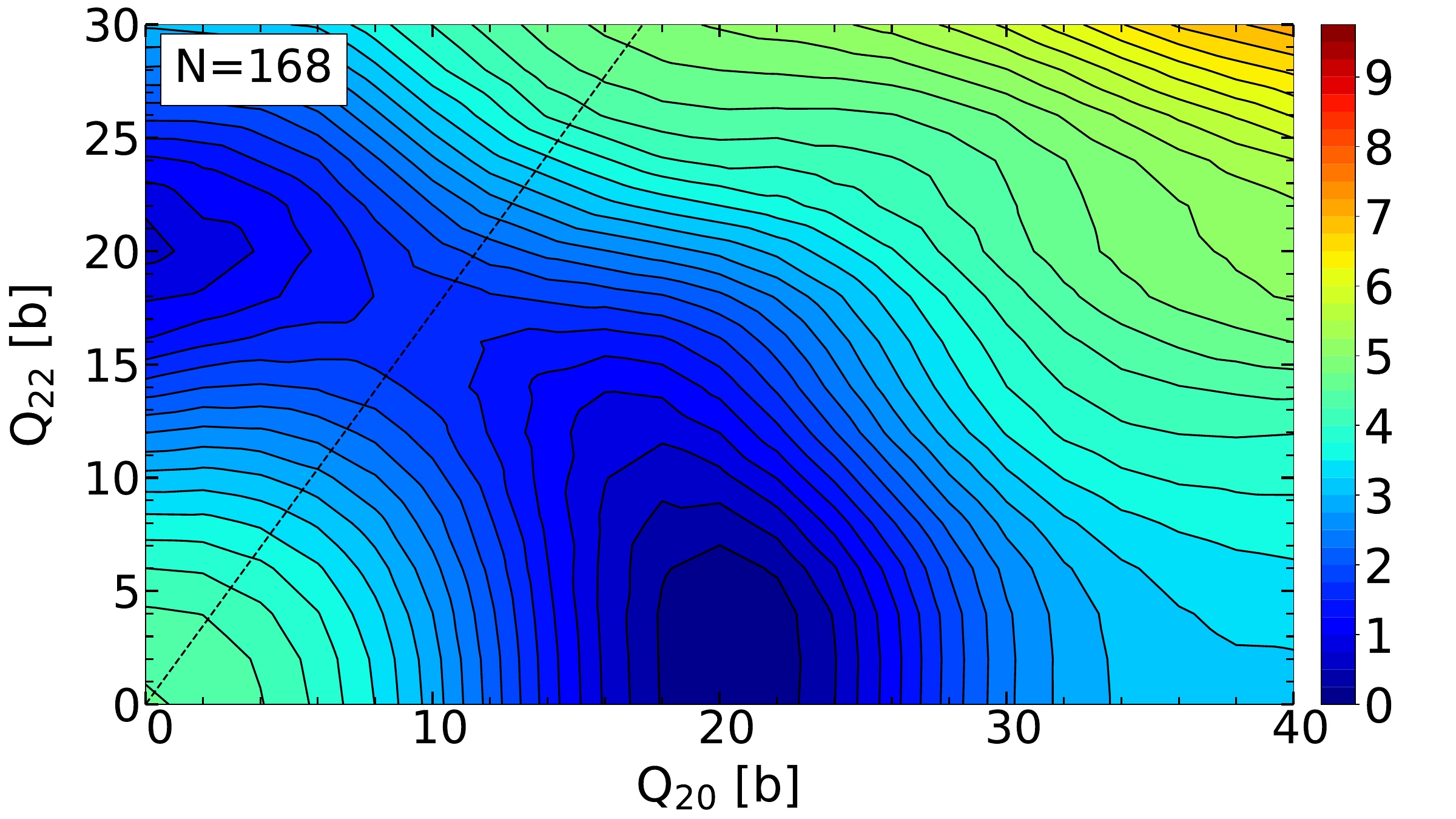}
    \includegraphics[width=0.48\linewidth]{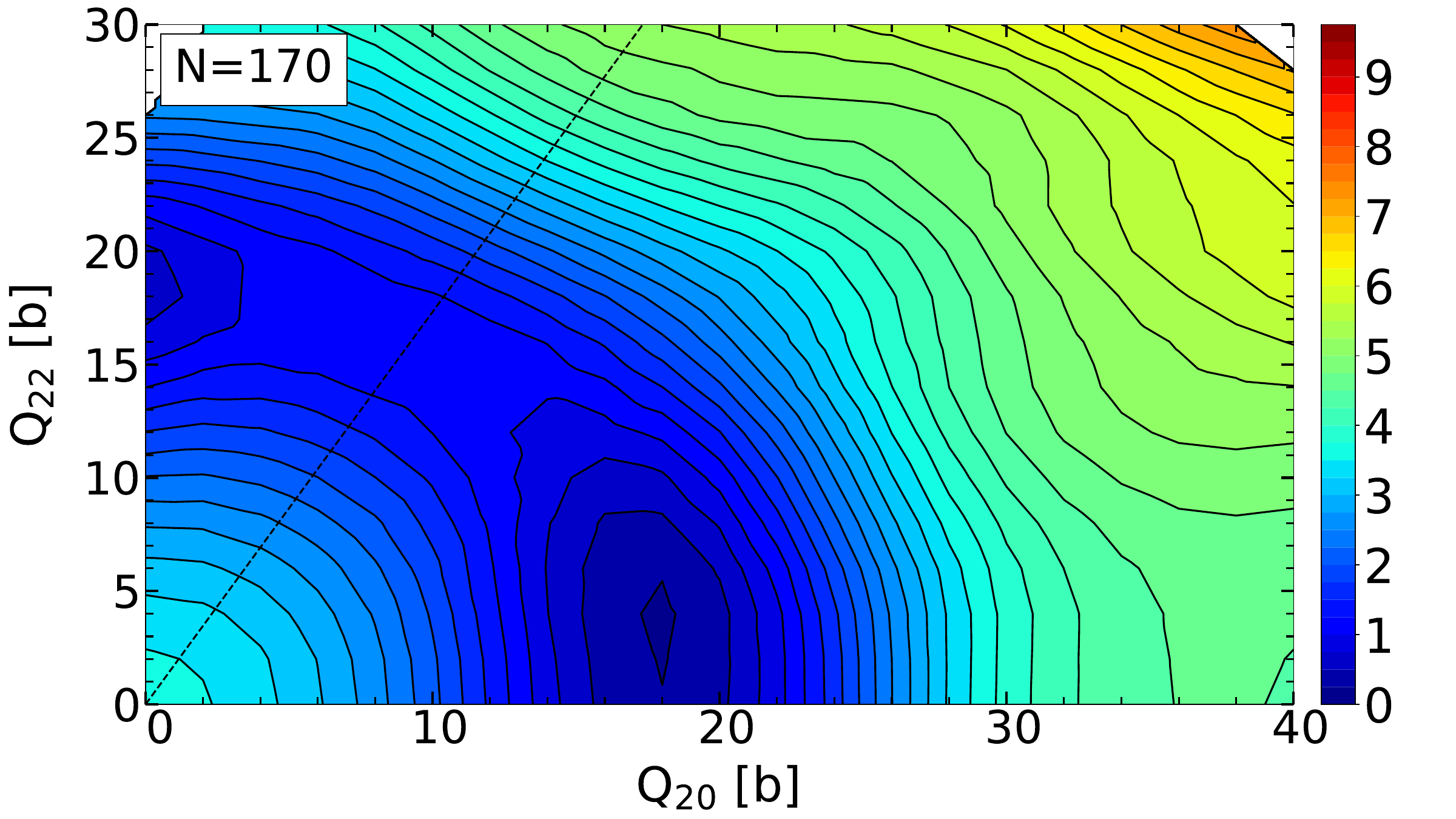}
    \includegraphics[width=0.48\linewidth]{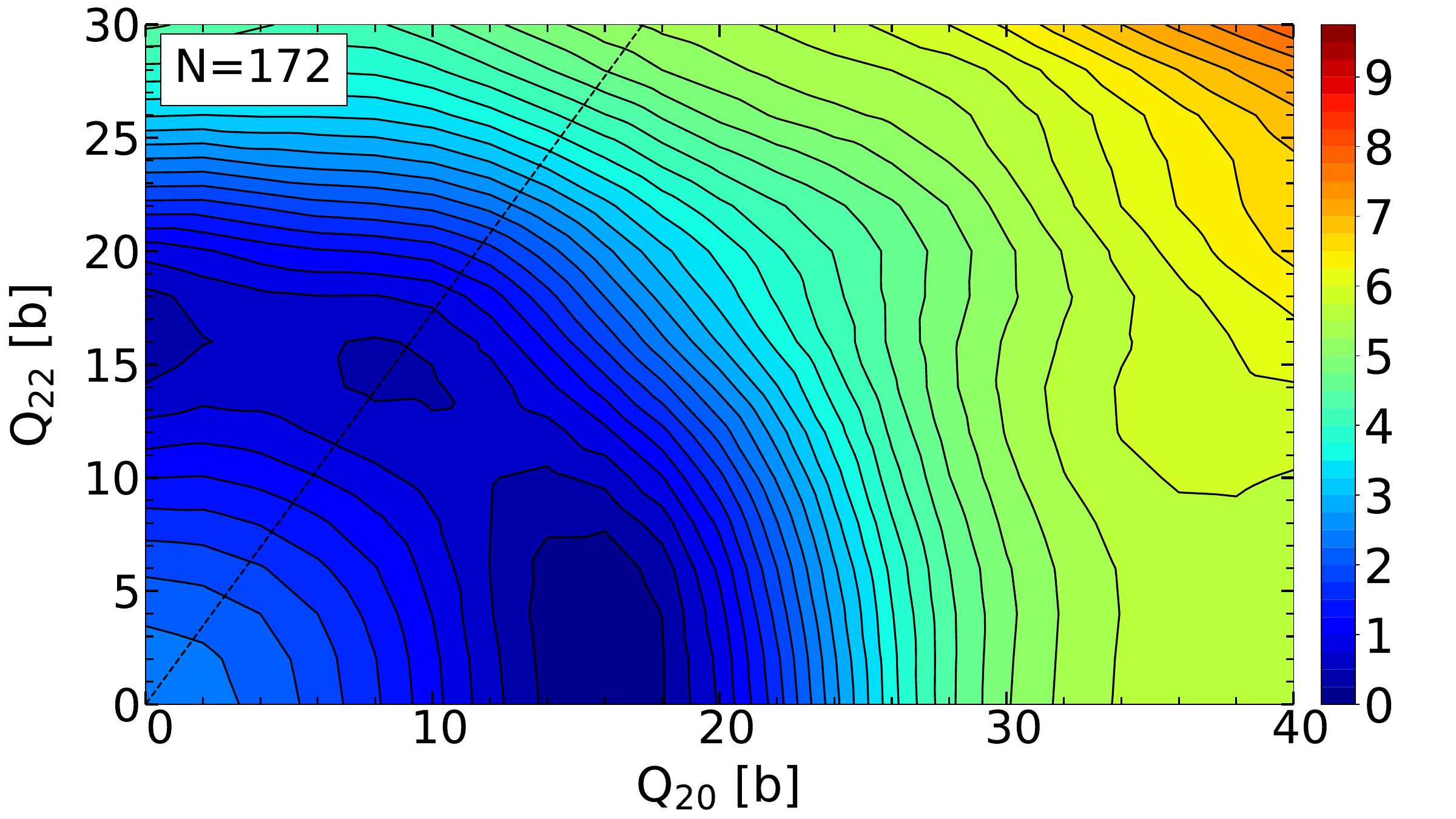}
    \includegraphics[width=0.48\linewidth]{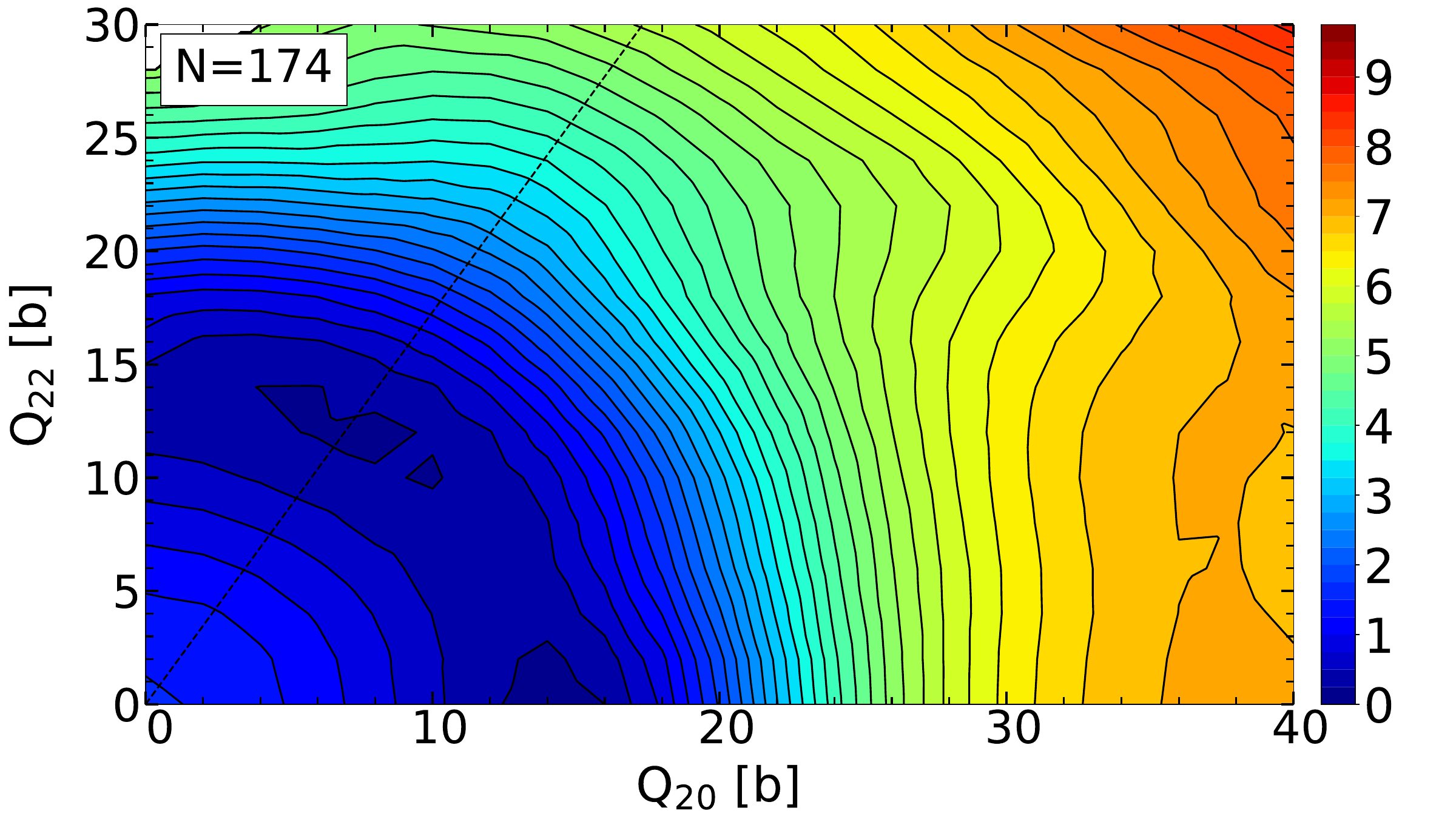}
    \caption{Potential energy surfaces in the $(Q_{20},Q_{22})$ plane 
            for plutonium isotopes around $N=170$.  
            The dashed line shows the line of the oblate shape. }
    \label{fig:Pu_isotopes_PES}
\end{figure*}

Figure \ref{fig:Pu_isotopes_PES} shows the potential energy surfaces of five 
neutron-rich plutonium isotopes in the $(Q_{20},Q_{22})$ plane. 
We employed the triaxial code {\sc{hfodd}} \cite{schunck2017solution} to perform constrained HFB 
calculations with the SLy4 parametrization and its corresponding pairing 
strengths listed in Table \ref{tab:pairing_strengths}. 
The 2D $(Q_{20},Q_{22})$ grid is $[0,40\;\mathrm{b}] \times [0,30\;\mathrm{b}]$ 
with steps of $\delta Q_{20} = \delta Q_{22} = 2$ b. 
In {\sc{hfodd}} all the quasiparticle wavefunctions 
were expanded in a deformed harmonic oscillator basis of $N = 20$ shells and 
quadrupole deformation $\beta_2 = 0.2$, with up to $N_{\rm states} = 1500$ 
states. The oscillator length was set automatically by
the {\sc{hfodd}} convention.

\input{supplement.bbl}

%% file: main.bbl
\providecommand{\noopsort}[1]{}

%% file: supplement.bbl
\providecommand{\noopsort}[1]{}